\documentclass[prd,aps,12pt,nofootinbib,superscriptaddress]{revtex4-1}

\usepackage{amsmath}
\usepackage[normalem]{ulem}

\usepackage{graphicx}

\usepackage{subfigure}

\newcommand{\beq}{\begin{equation}}
\newcommand{\eeq}{\end{equation}}

\newcommand{\ber}{\begin{eqnarray}}
\newcommand{\eer}{\end{eqnarray}}

\def\m{{\rm m}}

\def\C{{\cal C}}

\def\sss{\scriptscriptstyle}

\def\beb{{b}}

\def\0{{\scriptscriptstyle (0)}}
\def\1{{\scriptscriptstyle (1)}}

\def\l{\left}
\def\r{\right}

\def\O{{\cal O}}
\def\v{V}

\def\s{c_s^2\,}

\def\w{{\rm (bulk)}}
\def\wf{{\rm (brane)}}
\def \lleq {\lower0.9ex\hbox{ $\buildrel < \over \sim$} ~}
\def \ggeq {\lower0.9ex\hbox{ $\buildrel > \over \sim$} ~}

\begin{document}

\title{Cosmological perturbations on the Phantom brane}

\author{Satadru Bag}\email{satadru@iucaa.in}
\affiliation{Inter-University Centre for Astronomy and Astrophysics, Pune, India} %

\author{Alexander Viznyuk}\email{viznyuk@bitp.kiev.ua}
\affiliation{Bogolyubov Institute for Theoretical Physics, Kiev 03680, Ukraine} %
\affiliation{Department of Physics, Taras Shevchenko National University, Kiev, Ukraine} %

\author{Yuri Shtanov}\email{shtanov@bitp.kiev.ua}
\affiliation{Bogolyubov Institute for Theoretical Physics, Kiev 03680, Ukraine} %
\affiliation{Department of Physics, Taras Shevchenko National University, Kiev, Ukraine} %

\author{Varun Sahni}\email{varun@iucaa.in}
\affiliation{Inter-University Centre for Astronomy and Astrophysics, Pune, India} %

\begin{abstract}
We obtain a closed system of equations for scalar perturbations in a multi-component
braneworld. Our braneworld possesses a phantom-like equation of state at late times,
$w_{\rm eff} < -1$, but no big-rip future singularity. In addition to matter and
radiation, the braneworld possesses  a new effective degree of freedom -- the `Weyl
fluid' or `dark radiation'. Setting initial conditions on super-Hubble spatial scales at
the epoch of radiation domination, we evolve perturbations of radiation, pressureless
matter and the Weyl fluid until the present epoch. We observe a gradual decrease in the
amplitude of the Weyl-fluid perturbations after Hubble-radius crossing, which results in
a negligible effect of the Weyl fluid on the evolution of matter perturbations on spatial
scales relevant for structure formation. Consequently, the quasi-static approximation of
Koyama and Maartens provides a good fit to the exact results during the matter-dominated
epoch. We find that the late-time growth of density perturbations on the brane proceeds
at a faster rate than in $\Lambda$CDM\@. Additionally, the gravitational potentials
$\Phi$ and $\Psi$ evolve differently on the brane than in $\Lambda$CDM, for which $\Phi =
\Psi$. On the brane, by contrast, the ratio $\Phi/\Psi$ exceeds unity during the late
matter-dominated epoch ($z \lesssim 50$). These features emerge as {\em smoking gun\/}
tests of phantom brane cosmology and allow predictions of this scenario to be tested
against observations of galaxy clustering and large-scale structure.

\end{abstract}

\maketitle

\tableofcontents

\section{Introduction}\label{sec: intro}

Cosmology, during the past two decades, has witnessed the introduction and development of
several bold new theoretical ideas. One, especially radical paradigm, involves the
braneworld concept. According to this paradigm (see \cite{Maartens:2010ar} for a review),
our universe is a lower-dimensional hypersurface (the `brane') embedded in a
higher-dimensional spacetime (the `bulk'). A new feature of the braneworld paradigm,
which distinguishes it from the earlier Kaluza--Klein constructs, is that spacetime
dimensions orthogonal to the brane need not be compact but could be `large' \cite{ADD}
and even infinite in length. In the simplest and most thoroughly investigated
cosmological models, there is only one large extra dimension accessible to gravity, while
all standard-model fields are assumed to be trapped on the brane.  From the viewpoint of
our four-dimensional world, this manifests as a modification of gravity.  In the seminal
Randall--Sundrum (RS) model \cite{RS}, gravity is modified on relatively small spatial
scales.  Apart from other interesting applications, this model was used to provide an
alternative explanation of galactic rotation curves and X-ray profiles of galactic
clusters without invoking the notion of dark matter \cite{Rotation curves}.

An important class of braneworld models contains the so-called `induced-gravity' term in
the action for the brane (it is induced by quantum corrections from the matter fields,
hence the term), and modifies gravity on relatively large spatial scales. First proposed
in \cite{Collins:2000yb, DGP, Shtanov:2000vr}, it has become known as the
Dvali--Gabadadze--Porrati (DGP) model.  Depending on the embedding of the brane in the
bulk space, this model has two branches of cosmological solutions \cite{Deffayet:2000uy}.
The `self-accelerating' branch was proposed to describe cosmology with late-time
acceleration without bulk and brane cosmological constants \cite{Deffayet:2001pu}, while
the `normal' branch requires at least a cosmological constant on the brane (called
 brane
tension) to accelerate cosmic expansion.
The self-accelerating branch was later shown to
be plagued by the existence of ghost excitations \cite{Ghosts}.
Without any additional
modification, this leaves the normal branch as the only physically viable
solution of
this braneworld model, consistent with current cosmological observations
of cosmic acceleration.  It is this
braneworld model that will be the subject of investigation in this paper.

As a model of dark energy, the normal braneworld branch exhibits an interesting generic
feature of {\em super-acceleration\/} which is reflected in the phantom-like effective
equation of state $w_{\rm eff} < -1$ \cite{Sahni:2002dx, Alam_Sahni, Lue:2004za}. Interestingly, the
Phantom brane smoothly evolves to a de~Sitter stage without running into a `Big-Rip'
future singularity typical of conventional phantom models. Such a phantom-like equation
of state appears to be consistent with the most recent set of observations of type Ia
supernovae combined with other data sets \cite{Rest:2013mwz}. The Phantom brane has a
number of interesting properties: (i) it is ghost-free and is characterized by the
effective equation of state $w_{\rm eff} < -1$, (ii) for an appropriate choice of
cosmological parameters, even a spatially flat braneworld can `loiter'
\cite{Sahni:2004fb}, (iii) the Phantom brane possesses  the remarkable  property of
`cosmic mimicry,' wherein a high-density braneworld exhibits the
{\em precise\/} expansion history of $\Lambda$CDM \cite{Sahni:2005mc}; for reviews, see
\cite{Sahni:review}.  Just like the Randall--Sundrum model, this braneworld model was
also used as an alternative explanation of rotation curves in galaxies without dark
matter \cite{Viznyuk:2007ft}.

The structure of the universe on the largest scales is spectacular, and  consists of a
`cosmic web' of intertwining galactic superclusters separated from each other by large
voids. Whereas the full description of the supercluster--void complex demands a knowledge
of non-linear gravitational clustering, useful insight into structure formation can
already be gleaned from the linear (and weakly non-linear) approximation
\cite{Sahni-Coles}. Linearized gravitational clustering in the braneworld model
encounters obvious difficulties and complications connected with the existence of a large
extra dimension.  One has to take into account the corresponding dynamical degree of
freedom and specify appropriate boundary conditions in the bulk space.  In the simple
case of a spatially flat brane, the extra dimension is noncompact, and one has to deal
with its spatial infinity. The bulk gravitational effects then lead to a non-local
character of the resulting equations on the brane.  In spite of this difficulty, by using
a very convenient Mukohyama master variable and master equation \cite{Mukohyama:2000ui,
Mukohyama:2001yp}, some progress has been made in this direction \cite{Deffayet:2002fn,
Deffayet:2004xg, Koyama:2005kd, Koyama:2006ef, Sawicki:2006jj, Song:2007wd,
Cardoso:2007xc, Seahra:2010fj} by employing various plausible simplifying assumptions or
approximations and by direct numerical integration. Most successful amongst these has
been the quasi-static (QS) approximation due to Koyama and Maartens \cite{Koyama:2005kd}
which is based on the assumption of {\em slow temporal evolution\/} of (all)
five-dimensional perturbations on sub-Hubble spatial scales, when compared with spatial
gradients (for an extension into the non-linear regime, see \cite{non-linear}). The
behavior of perturbations on super-Hubble spatial scales was investigated within the
scaling ansatz proposed in \cite{Sawicki:2006jj} and further developed in
\cite{Song:2007wd, Seahra:2010fj}. The validity of the quasi-static and scaling
approximations was confirmed by numerical integration of the perturbation equations in
five dimensions
\cite{Cardoso:2007xc, Seahra:2010fj}. %It is important to note here that these
%semi-analytic and numerical methods were based on the assumption that bulk perturbations
%are generated causally from the brane perturbations, so that the Mukohyama master
%variable vanishes on the past Cauchy horizon in the bulk.

In our previous work \cite{Viznyuk:2013ywa}, we addressed the problem of scalar
cosmological perturbations in a matter-dominated braneworld model. We considered a
marginally spatially closed (with topology $S^3$) braneworld model, in which the
`no-boundary' smoothness conditions for the five-dimensional perturbations were set in
the four-ball bulk space bounded by the $S^3$ brane.  In the limit when the spatial
curvature radius of the brane was large, we were able to arrive at a {\em closed\/}
system of equations for scalar cosmological perturbations without any simplifying
assumptions.

Our approach differs from the semi-analytic theory and numerical
computations developed in \cite{Sawicki:2006jj, Song:2007wd, Cardoso:2007xc,
Seahra:2010fj}. First of all, we only require the regularity condition in the compact
bulk space but do not impose any additional boundary conditions in the bulk; in
particular, we do not demand the bulk perturbations to vanish on the past Cauchy horizon
of the brane.  At the same time, as we have shown, due to the same regularity condition,
the system of equations for perturbations becomes effectively closed on the brane and
does not require integration in the bulk. This is a great simplification of the theory.

In our approach, the dynamical entities that describe perturbations on the brane are the
usual matter components and the so-called Weyl fluid, or dark radiation, which stems from
the projection of the five-dimensional Weyl tensor onto the brane.  The closed system of
dynamical equations allows one to trace the behavior of matter and Weyl-fluid
perturbations once the initial conditions for these quantities are specified. In
particular, for modes well inside the Hubble radius during matter domination, the matter
density perturbation $\delta_m = \delta \rho_m / \rho_m$ evolves as [see also
Eq.~\eqref{Delta matter domination g-r solution} below]
\beq\label{Delta matter introduction}
\delta_m  = {\cal M} a + \frac{W_1}{a^{5/4}} \cos \frac{\sqrt{2} k}{a H} +
\frac{W_2}{a^{5/4}} \sin \frac{\sqrt{2} k}{a H} \, ,
\eeq
where $a$ is the scale factor, and ${\cal M}$, $W_1$, $W_2$ are integration constants.
Apart from the usual growing mode ${\cal M} a$, we observe two oscillating modes with
decreasing amplitudes. These modes are induced by the dynamics of the Weyl fluid, or dark
radiation. Note that, since these oscillatory modes have their origin in the
bulk, they are absent in the scaling approximation of \cite{Sawicki:2006jj, Song:2007wd,
Seahra:2010fj} or in the quasi-static approximation of \cite{Koyama:2005kd}.

Whether or not the presence of such extra modes, with dynamical origin in the
bulk, can be significant for the braneworld model, depends upon the amplitudes $W_1$ and
$W_2$, and these, in turn, are determined by the primordial power spectrum of the Weyl
fluid and its evolution during the radiation-dominated epoch.  This calls for a
development of the theory of scalar cosmological perturbations for a universe filled with
several components, each with an arbitrary equation of state. Such a treatment will
enable one to follow the evolution of perturbations starting from deep within the
radiation-dominated regime all the way up to the current stage of accelerated expansion.
This will be the main focus of the present paper.

Our paper is organized as follows. In the next section, we describe the background
cosmological evolution of the normal branch of the braneworld model embedded in a flat
five-dimensional bulk. In Sec.~\ref{sec: Scalar cosmological perturbations on the brane},
we investigate the system of equations describing scalar cosmological perturbations in
this model.  This system is not closed on the brane because the evolution equation for
the anisotropic stress from the bulk degree of freedom projected to the brane (the
so-called Weyl fluid, or dark radiation) is missing.  We solve this problem by proceeding
to a {\em marginally\/} closed braneworld and imposing the regularity conditions in the
bulk, as described in \cite{Viznyuk:2013ywa}. The system of equations on the brane (in
the limit of a large spatial radius) now becomes closed and can therefore be used for the
analysis of perturbations.  In Sec.~\ref{sec: Weyl perturbations at the
radiation-dominated epoch}, we discuss possible ways of setting initial conditions for
Weyl-fluid perturbations and investigate the evolution of all perturbations starting from
super-Hubble scales until the end of the radiation-dominated epoch. Perturbations during
matter-domination are considered in Sec.~\ref{sec: Weyl perturbations in the
matter-dominated epoch}\@. In Sec.~\ref{sec: Numerical integration}, we present the
results of numerical integration of the joint system of equations describing
perturbations in radiation, pressureless matter and the Weyl fluid and compare these with
$\Lambda$CDM and with the results from the quasi-static approximation.  Our results are
summarized in Sec.~\ref{sec: conclusion}.

\section{Background cosmological evolution}\label{sec: background}

Our braneworld model has the action \cite{Collins:2000yb, Shtanov:2000vr,
Sahni:2002dx}
\beq \label{action}
S = M^3 \left[\int_{\rm bulk} \left( {\cal R} - 2 \Lambda \right) - 2 \int_{\rm brane} K
\right] +  \int_{\rm brane} \left( m^2 R - 2 \sigma \right) + \int_{\rm brane} L \left(
g_{\mu\nu}, \phi \right) \, ,
\eeq
where ${\cal R}$ is the scalar curvature of the five-dimensional bulk, and $R$ is the
scalar curvature corresponding to the induced metric $g_{\mu\nu}$ on the brane. The
symbol $L \left( g_{\mu\nu}, \phi \right)$ denotes the Lagrangian density of the
four-dimensional matter fields $\phi$ whose dynamics is restricted to the brane so that
they interact only with the induced metric $g_{\mu\nu}$. The quantity $K$ is the trace of
the symmetric tensor of extrinsic curvature of the brane. All integrations over the bulk
and brane are taken with the corresponding natural volume elements. The universal
constants $M$ and $m$ play the role of the five-dimensional and four-dimensional Planck
masses, respectively. The symbol $\Lambda$ denotes the bulk cosmological constant, and
$\sigma$ is the brane tension.

Action \eqref{action} leads to the following effective equation on the brane
\cite{Shiromizu:1999wj, Sahni:2005mc}:
\begin{equation}
G_{\mu\nu} + \left(\frac{\Lambda_{\rm RS}}{\beb + 1}\right) g_{\mu\nu}
= \left(\frac{\beb}{\beb + 1}\right) \frac{1}{m^2} \, T_{\mu\nu} +
\left(\frac{1}{\beb + 1}\right) \left[ \frac{1}{M^6} Q_{\mu\nu} -
\C_{\mu\nu} \right] \, ,\label{effective}
\end{equation}
where
\begin{equation}\label{beta}
\beb = \frac{\sigma\ell}{3M^3}\,, \qquad
\ell=\frac{2m^2}{M^3}\,,\qquad \Lambda_{\rm
RS}=\frac{\Lambda}{2}+\frac{\sigma^2}{3M^6}
\end{equation}
are convenient parameters, and
\begin{equation}\label{q}
Q_{\mu\nu} = \frac13 E E_{\mu\nu} - E_{\mu\lambda} E^{\lambda}{}_\nu + \frac12
\left(E_{\rho\lambda} E^{\rho\lambda} - \frac13 E^2 \right) g_{\mu\nu}\,,
\end{equation}
\begin{equation}\label{bare einstein}
E_{\mu\nu} \equiv m^2 G_{\mu\nu} - T_{\mu\nu} \,,\qquad E=E^{\mu}{}_\mu\,.
\end{equation}

Gravitational dynamics on the brane is not closed because of the presence of the
symmetric traceless tensor $\C_{\mu\nu}$ in \eqref{effective}, which stems from the
projection of the five-dimensional Weyl tensor from the bulk onto the brane. We are free
to interpret this tensor as the stress-energy tensor of some effective fluid, which we
call the `Weyl fluid' in this article (in some works, the term `dark radiation' is also
used).

The tensor $\C_{\mu\nu}$
is not freely specifiable on the brane, but is related to the tensor $Q_{\mu\nu}$ through
the conservation equation
\begin{equation}\label{conserv_weyl}
\nabla^\mu \left( Q_{\mu\nu} - M^6 \C_{\mu\nu} \right) = 0 \, ,
\end{equation}
which is a consequence of the Bianchi identity applied to \eqref{effective} and the law
of stress--energy conservation for matter:
\begin{equation}\label{conserv}
\nabla^\mu T_{\mu\nu} = 0 \, .
\end{equation}
In a universe consisting of several non-interacting components, the conservation law
\eqref{conserv} is satisfied by each component separately.

The cosmological evolution of the Friedmann--Robertson--Walker (FRW) brane
\begin{equation}
\label{FRW} ds^2 = - d t^2 + a^2 (t) \gamma_{ij} d x^i dx^j \,
\end{equation}
can be obtained from \eqref{effective} with the following result \cite{Collins:2000yb,
Shtanov:2000vr, Deffayet:2000uy, Deffayet:2001pu, Sahni:2002dx}:
\begin{equation} \label{background}
H^2 + \frac{\kappa}{a^2} = \frac{\rho + \sigma}{3 m^2} + \frac{2}{\ell^2}\left[ 1 \pm
\sqrt{1 + \ell^2 \left(\frac{\rho + \sigma}{3 m^2} - \frac{\Lambda}{6} - \frac{C}{a^4}
\right)} \right] \, .
\end{equation}
Here, $H\equiv \dot{a}/a$ is the Hubble parameter, $\rho=\rho(t)$ is the energy density of
matter on the brane and $C$ is a constant resulting from the presence of the symmetric
traceless tensor $\C_{\mu\nu}$ in the field equations \eqref{effective}. The parameter
$\kappa = 0, \pm 1$ corresponds to different spatial geometries of the maximally
symmetric spatial metric $\gamma_{ij}$.

The sign ambiguity in front of the square root in equation \eqref{background} reflects
the two different ways in which the bulk can be bounded by the brane
\cite{Deffayet:2000uy, Sahni:2002dx}, resulting in two different branches of solutions.
These are usually called the normal branch (lower sign) and the self-accelerating branch
(upper sign).

In what follows, we investigate the evolution of perturbations
in a marginally flat ($a H~\gg~1$) normal branch of the braneworld model embedded in flat
bulk spacetime (which means $C=0$, $\Lambda=0$). This setup will enable us to obtain a
closed system of equations on the brane. The background cosmological equation
(\ref{background}) then reduces to
\begin{equation} \label{background-flat}
H^2 = \frac{\rho + \sigma}{3 m^2} + \frac{2}{\ell^2}\left[ 1 -
\sqrt{1 + \ell^2 \left(\frac{\rho + \sigma}{3 m^2} \right)} \right]
=  \frac{1}{\ell^2} \left[ 1 - \sqrt{1 + \ell^2 \left(\frac{\rho +
\sigma}{3 m^2} \right)} \right]^2 \, ,
\end{equation}
or, equivalently:
\beq \label{brane Hubble}
\ell H = \,\sqrt{1 + \ell^2 \l(\frac{\rho + \sigma}{3 m^2}\r)}-1\,.
\eeq
One immediately sees that, in the regime $H\gg \ell^{-1}$, our braneworld expands like
$\Lambda$CDM with the gravitational constant $8 \pi G = 1/m^2$ and with the combination
$\sigma / m^2$ playing the role of the cosmological constant.

Our braneworld is assumed to be filled with a multi-component fluid, with total
energy density $\rho=\sum_\lambda \rho_\lambda$ and pressure $p =\sum_\lambda p_\lambda$.
The usual conservation law holds for each component separately:
\beq \label{GR-brane conservation}
\dot{\rho}_\lambda+3H \rho_\lambda(1 + w_\lambda)=0 \,,
\eeq
where $w_\lambda = p_\lambda/\rho_\lambda$ is the equation of state parameter for the
component labeled by $\lambda$.

In terms of the cosmological parameters
\beq
\Omega_\lambda = \frac{\rho_{\lambda 0}}{3m^2H_0^2} \, , \qquad \Omega_\sigma =
\frac{\sigma}{3m^2H_0^2} \, , \qquad \Omega_\ell = \frac{1}{\ell^2H_0^2} \, ,
\label{eq:cosmo_parameters}
\eeq
where $\ell = 2m^2/M^3$ and $H_0$ is the present value of the Hubble parameter, one can write the evolution
equation \eqref{brane Hubble} in the form
\beq \label{brane Hubble cosm param}
h(z)= \frac {H(z)}{H_0} = \sqrt{\sum_\lambda \Omega_\lambda (1 + z)^{3(1 + w_\lambda)} +
\Omega_\sigma + \Omega_\ell}\, - \sqrt{\Omega_\ell}\, ,
\eeq
where $1 + z = a_0 / a$, and $a_0$ is the present value of the scale factor. The
cosmological parameters are related through the equation
\begin{equation}\label{eq:cos_par}
 \Omega_{\sigma}=1+2\sqrt{\Omega_{\ell}}-\sum_\lambda \Omega_{\lambda} \, .
\end{equation}

For further convenience, we introduce the time-dependent parameters $\beta$ and
$\gamma$\,:
\beq \label{beta-3} \beta = {} - 2\, \sqrt{1 + \ell^2 \left(\frac{\sum_\lambda \rho_\lambda +
\sigma}{3 m^2} \right) } = {} - 2\,(1+\ell H)\, ,
\eeq
\begin{equation} \label{gamma-flat}
3\,\gamma - 1 =\, \frac{\dot \beta}{H \beta} \,=\, \frac{\ell
\dot{H}}{H (1 + \ell H)}  \, .
\end{equation}
Then, from \eqref{brane Hubble} and \eqref{GR-brane conservation} one can derive a useful
equation
\beq \label{brane dot H} \dot{H} = {} - \l(1+\frac{2}{\beta}\r) \frac{\sum_\lambda (\rho_\lambda + p_\lambda)}{2m^2}
\, .
\eeq

Restricting our attention for the moment to the present epoch, when the density of matter
greatly exceeds that of radiation, we find
\beq
 h(z) = \sqrt{\Omega_{m}(1+z)^3+\Omega_{\sigma}+\Omega_{\ell}}-\sqrt{\Omega_{\ell}} \, ,
\label{eq:h}
\end{equation}
where
\begin{equation}\label{eq:closure}
\Omega_{\sigma}=1+2\sqrt{\Omega_{\ell}}-{\Omega_{m}}\;,
\end{equation}
so that
\ber
\beta &=& -\frac{2}{\sqrt{\Omega_{\ell}}}\sqrt{\Omega_{m}(1+z)^3+\Omega_{\sigma}+\Omega_{\ell}}\, , \\
3\gamma - 1 &=&
-\frac{\frac{3}{2}\Omega_{m}(1+z)^3}{\Omega_{m}(1+z)^3+\Omega_{\sigma}+\Omega_{\ell}} \,
.
\eer

\begin{figure}[hbt]
\centering
\includegraphics[width=.55\textwidth]{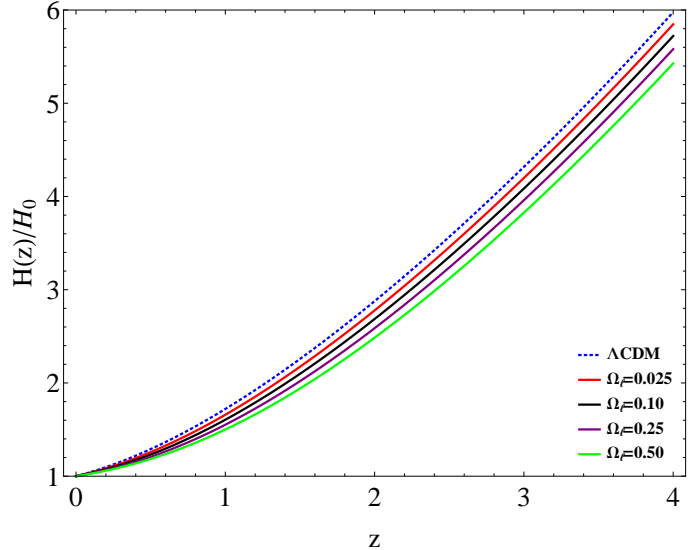}
\caption{
The expansion rate in $\Lambda$CDM (top/dotted blue) is compared to that in the Phantom brane with
$\Omega_{\ell}=0.025$, $0.1$, $0.25$, $0.5$ (top to bottom/solid), and $\Omega_m=0.28$.
One finds that past expansion is slower in the Phantom brane than in $\Lambda$CDM.}
 \label{fig:expansion}
\end{figure}

An important feature of the Phantom-brane is that, for a given value of $\Omega_{m}$, its
expansion rate is {\em slower\/} than that in $\Lambda$CDM, see
figure~\ref{fig:expansion}. This property of our model is of special significance when
one compares its observational predictions with observational data \cite{Sahni:2014ooa}.
Indeed, recent measurements of the expansion rate at high redshifts using the data on
baryon acoustic oscillations (BAO) indicate $H(z) = 222 \pm 7$ km/sec/Mpc at $z = 2.34$,
which is below the value predicted by $\Lambda$CDM \cite{bao_2014}. This tension is
alleviated in the Phantom brane \cite{Sahni:2014ooa}.

The {\em effective equation of state\/} (EOS) of dark energy on the brane is given by
\cite{ss06}
\begin{equation}\label{eq:weff}
 w(z)=\dfrac{\dfrac{2}{3}(1+z)\dfrac{h'}{h}-1}{1-\Omega_{m} (1+z)^3 /h^2}\;,
\end{equation}
where the prime denotes differentiation with respect to $z$. Substituting (\ref{eq:h})
into (\ref{eq:weff}) one finds that the present value of the EOS is
\begin{equation}\label{eq:weff_bw_0}
 w_0 \equiv w(z=0)=-1-\dfrac{\Omega_{m}}{1-\Omega_{m}} \left(\dfrac{\sqrt{\Omega_{\ell}}}{1+\sqrt{\Omega_{\ell}}}\right)\;.
\end{equation}
Consequently, the Phantom brane possesses a phantom EOS $w_0 < -1$, just as its name
suggests.

%%%%%%%%%%%%%%%%%%%%%%%%%%%%%%%%%%%%%%%%%%%%%%%%%%%%%%%%%%%%%%%%%%%%%%%%%%%%%%%%%%%%%5
\section{Evolution of scalar cosmological perturbations}
\label{sec: Scalar cosmological perturbations on the brane}

\subsection{General equations for a multi-component fluid}
\label{Sec: Multi-component fluid on the brane}

Scalar cosmological perturbations of the induced metric on the brane are most
conveniently described by the relativistic potentials $\Phi$ and $\Psi$ in the
longitudinal gauge:
\begin{equation} \label{metric}
ds^2 = - ( 1 + 2 \Phi) d t^2 + a^2 ( 1 - 2 \Psi) \gamma_{ij} d x^i dx^j \, .
\end{equation}

The components of the linearly perturbed stress--energy tensor of the $\lambda$-component
of matter
can be parameterized as follows:\footnote{The spatial indices $i,j,\ldots$ in purely
spatially defined quantities (such as $v_i$ and $\delta \pi_{ij}$) are always raised and
lowered using the spatial metric $\gamma_{ij}\,$; in particular, $\gamma^i{}_j =
\delta^i{}_j$. The symbol $\nabla_i$ denotes the covariant derivative with respect to the
spatial metric $\gamma_{ij}$, and the spatial Laplacian is $\nabla^2 = \nabla^i
\nabla_i$.}
\begin{equation}
\delta T_{(\lambda)}{}^{\mu}{}_\nu = \left(
\begin{array}{cc}
\displaystyle - \delta \rho_{\lambda} \, , & \displaystyle - \nabla_i v_{\lambda}
\medskip \\
\displaystyle  \frac{\nabla^i v_{\lambda}}{a^2} \, , & \displaystyle \delta
p_{\lambda} \delta^i{}_j + \frac{\zeta_\lambda{}^{i}{}_j}{a^2}
\end{array}
\right) \, ,
\end{equation}
where $\delta \rho_{\lambda}$, $\delta p_{\lambda}$, $v_{\lambda}$, and
$\zeta_{\lambda\,ij} = \left( \nabla_i \nabla_j - \frac13 \gamma_{ij} \nabla^2 \right)
\zeta_{\lambda}$ describe scalar perturbations.

Although the term $\rho_\C=C/a^4$ [see \eqref{background}], which can be treated as the
homogeneous part of the Weyl fluid, or dark radiation, was set to zero in
\eqref{background-flat}, perturbations of the Weyl fluid should be taken into
account. Therefore, in a similar way we introduce scalar perturbations $\delta
\rho_\C$, $v_\C$, and $\delta \pi_\C$ of the traceless tensor $\C_{\mu\nu}$:
\begin{equation} \label{weyl projection definition}
m^2 \delta \C^\mu{}_\nu = \left(
\begin{array}{cc}
\displaystyle - \delta \rho_\C
\, , & \displaystyle - \nabla_i v_\C  \medskip \\
\displaystyle \frac{\nabla^i v_\C}{a^2} \, , & \displaystyle \frac{\delta
\rho_\C}{3 } \delta^i{}_j + \frac{\delta \pi^i{}_j}{a^2}
\end{array}
\right) \, ,
\end{equation}
where $\delta \pi_{ij} = \left( \nabla_i \nabla_j - \frac13
\gamma_{ij} \nabla^2 \right) \delta \pi_\C$.

We call $v_{\lambda}$ and $v_\C$ the momentum potentials for the matter components and
Weyl fluid, respectively; the quantities $\delta \rho_{\lambda}$ and $\delta \rho_\C$ are
their energy-density perturbations, and $\zeta_{\lambda}$ and $\delta\pi_\C$ are the
scalar potentials for their anisotropic stresses.

In this notation, the effective field equation leads to the following system for
perturbations in the Fourier-space representation with respect to comoving spatial
coordinates \cite{Viznyuk:2013ywa, Shtanov:2007dh, Viznyuk:2012oda}:
\begin{equation}
-\, \frac{k^2}{a^2}\, \Psi = \left(1 + \frac{2}{\beta} \right) \frac{\sum_{\lambda}(\delta
\rho_\lambda + 3 H v_\lambda)}{2 m^2} + \frac{\left(\delta \rho_\C + 3 H v_\C
\right)}{m^2 \beta } \, , \label{con-delta1}
\end{equation}
\begin{equation}
\dot \Psi + H \Phi = \left(1 + \frac{2}{\beta} \right) \frac{\sum_{\lambda}
v_\lambda}{2m^2} + \frac{v_\C}{m^2\beta} \, , \label{con-v1}
\end{equation}
\begin{equation}
\Psi-\Phi = \frac{4\,\delta \pi_\C}{m^2\beta (1+3 \gamma)} \, ,
\label{pi1}
\end{equation}
where $k$ is the comoving wavenumber and
$\beta$, $\gamma$ were defined in (\ref{beta-3}), (\ref{gamma-flat}).
From the conservation law \eqref{conserv} applied
to each component separately, we have
\begin{equation}
\delta \dot{\rho}_\lambda + 3 H ( \delta \rho_\lambda + \delta p_\lambda ) = {} -
\frac{k^2}{a^2}\,v_\lambda + 3 (\rho_\lambda + p_\lambda ) \dot \Psi \, , \label{brane
perturb matter conserv rho}
\end{equation}
\begin{equation}
\dot{v}_\lambda + 3 H v_\lambda = \delta p_\lambda + (\rho_\lambda + p_\lambda) \Phi \, . \label{brane
perturb matter conserv v}
\end{equation}

For simplicity, we have assumed here that the anisotropic stresses of all matter
components vanish: $\zeta_\lambda = 0$. Note that
in general relativity these assumptions would
lead to the equality of relativistic gravitational potentials $\Phi$ and $\Psi$, but, in
the braneworld, this is not the case due to the presence of the anisotropic stress
of the Weyl fluid $\delta\pi_\C$ [see \eqref{pi1}]. The contribution from $\delta\pi_\C$
in the braneworld model cannot be ignored, but should be established from the analysis of
five-dimensional perturbations in the bulk.

The pressure perturbations $\delta p_\lambda$ are usually decomposed into adiabatic
and isentropic parts:
\beq \label{delta p decomposition}
\delta p_\lambda = c_{s \lambda}^2 \delta\rho_\lambda + (\rho_\lambda + p_\lambda) \Gamma_\lambda \,,
\eeq
where
\beq \label{sound speed def}
c_{s \lambda}^2 \equiv \frac{\dot{p}_\lambda}{\dot{\rho}_\lambda}=w_\lambda-\frac{\dot{w}_\lambda}{3H(1+w_\lambda)}
\eeq
is the adiabatic sound speed, and $\Gamma_\lambda$ describe the non-adiabatic pressure
perturbations. In what follows, we restrict ourselves to
 adiabatic perturbations by setting
$\Gamma_\lambda = 0$. At the same time, the equation of state parameters $w_\lambda$ can
be arbitrary functions of time.

Equation \eqref{conserv_weyl} serves as a conservation law for the perturbation of the
Weyl fluid:
\begin{equation}
\delta \dot \rho_\C + 4 H \delta \rho_\C = - \,\frac{k^2}{a^2}
\,v_\C \, , \label{w11}
\end{equation}
\beq
\dot v_\C + 3 H v_\C = \frac13 \delta \rho_\C + \frac{\beta (1 - 3 \gamma)}{6}
\sum_{\lambda}(\delta \rho_\lambda + 3 H v_\lambda) + \frac{m^2 \beta k^2}{3 a^2} \left[
\Phi - 3 \gamma \Psi \right]\, .  \label{w33}
\eeq

In the quasi-static approximation proposed by Koyama and Maartens in
\cite{Koyama:2005kd}, there arises the following approximate relation between $\delta
\pi_{\C}$ and $\delta \rho_{\C}$:
\beq \label{boundary conditions qs}
\delta \pi_{\C} \approx \frac{a^2}{2k^2}\,\delta \rho_{\C} \,.
\eeq
Relation \eqref{boundary conditions qs} is properly justified on sub-Hubble scales, where
$k \gg a H$. As noted in Sec.~\ref{sec: intro}, the Koyama-Maartens
relation cannot be used to study
the behavior of perturbations of the Weyl fluid during the early
radiative epoch on super-Hubble spatial scales.

A more general relation between $\delta\pi_\C$ and $\delta\rho_\C$ was derived in
\cite{Viznyuk:2013ywa} in the limit of a marginally closed braneworld:
\beq \label{boundary conditions no-boundary}
\delta \pi_{\C} = \frac{3a^4}{2k^4} \left[\delta {\ddot \rho}_{\C} +
\left(9 H - \frac{\dot{H}}{H}\right)\, \delta {\dot \rho}_{\C} +
\left(20 H^2 + \frac{k^2}{3a^2}\right)\, \delta \rho_{\C} \right] \,.
\eeq
Relation \eqref{boundary conditions no-boundary} accounts for the temporal evolution of
the Weyl fluid, which makes it possible to trace the evolution of perturbations right
from their initial values on super-Hubble spatial scales all the way until the present
time. We shall use equation \eqref{boundary conditions no-boundary} in the present
analysis.

To investigate perturbations of a multi-component fluid, we introduce convenient variables
\beq\label{dimensionless variables}
\delta_\lambda \equiv \frac{\delta\rho_\lambda}{\rho_\lambda+p_\lambda}\,, \qquad
\v_\lambda\equiv\frac{v_\lambda}{\rho_\lambda+p_\lambda}\,.
\eeq
The variables $\v_\lambda$ are proportional to the physical velocity potentials
$\v_\lambda=\v^{\rm phys}_\lambda/a$. It is reasonable to introduce similar variables for
the Weyl fluid.  The Weyl fluid, being described by a traceless effective stress-energy
tensor $\C_{\mu\nu}$ in (\ref{effective}), behaves in a way rather similar to radiation.
Hence, since the background density of the Weyl fluid vanishes, we use the radiation
background component to define
\beq\label{dimensionless variables Weyl}
\delta_\C \equiv \frac{\delta\rho_\C}{\rho_r+p_r}=\frac{3\delta\rho_\C}{4\rho_r}\,,
\qquad \v_\C\equiv\frac{v_\C}{\rho_r+p_r}=\frac{3v_\C}{4\rho_r}\, ,
\eeq
where $\rho_r$ and $p_r=\rho_r/3$ are, respectively, the energy density and pressure of
radiation. Naturally
this assumes the presence of radiation during all stages
of cosmological expansion, which is certainly true for our universe after inflation.

In terms of the new variables \eqref{dimensionless variables}, \eqref{dimensionless variables
Weyl}, we have the following closed system of equations on the brane:
\beq \label{GR Einstein Psi brane multi-component}
- \frac{k^2}{a^2}\,\Psi =
\l(1+\frac{2}{\beta}\r)\frac{\sum_{\lambda}(\rho_\lambda+p_\lambda)\l(\delta_\lambda + 3H
\v_\lambda\r)}{2m^2} + \frac{4 \rho_r\left(\delta_\C + 3 H \v_\C \right)}{3 m^2 \beta
}\,,
\eeq
\beq \label{GR Einstein upsilon brane multi-component}
\dot{\Psi} + H\Phi = \l(1 +
\frac{2}{\beta}\r)\frac{\sum_{\lambda}(\rho_\lambda+p_\lambda)\v_\lambda}{2m^2} + \frac{4
\rho_r\v_\C}{3 m^2\beta}\,,
\eeq
\beq \label{GR Einstein zeta brane multi-component}
\Psi-\Phi = \frac{4\,\delta \pi_\C}{m^2\beta (1+3 \gamma)}\,,
\eeq
\beq \label{GR conserv rho brane multi-component}
\dot{\delta}_\lambda = - \frac{k^2}{a^2}\,\v_\lambda + 3\dot{\Psi} \,,
\eeq
\beq \label{GR conserv upsilon brane multi-component}
\dot{\v}_\lambda -3H c_{s \sss \lambda}^2 \v_\lambda = c_{s \sss \lambda}^2\delta_\lambda
+ \Phi \, ,
\eeq
\begin{equation}
 \dot \delta_\C = - \,\frac{k^2}{a^2}
\,\v_\C \, , \label{rho C-1}
\end{equation}
\begin{equation}
\dot \v_\C - 3\gamma H \v_\C = \gamma \delta_\C +\frac{3\gamma
- 1}{4\rho_r} \sum_{\lambda}(\rho_\lambda+p_\lambda)\l(\delta_\lambda + 3H \v_\lambda\r) \,-
\frac{ k^2}{\rho_r a^2 (1 + 3 \gamma)} \, \delta \pi_\C \, , \label{v C}
\end{equation}
\beq \label{no-boundary dim less variable}
\delta \pi_{\C} = \frac{2 \rho_r a^4}{k^4} \left[ {\ddot \delta}_{\C} +
\left(1 - \frac{\dot{H}}{H^2}\right)\,H {\dot \delta}_{\C} +
 \frac{k^2}{3a^2}\, \delta_{\C} \right] \,.
\eeq

Remarkably, equations \eqref{rho C-1}--\eqref{no-boundary dim less variable} lead to a
single equation for the variable $\delta_\C$:
\beq\label{delta C multicomponent}
\ddot\delta_\C + \l(\frac{2\beta}{2+\beta} - 3\gamma\r)H\dot\delta_\C + \frac{k^2}{3 a^2}(2+3\gamma)\delta_\C
= - \frac{k^2 (1+3\gamma)}{4\rho_r a^2}\sum_{\lambda}(\rho_\lambda+p_\lambda)\Delta_\lambda\,,
\eeq
where
\beq\label{gauge inv variable delta}
\Delta_\lambda \equiv
\delta_\lambda+3 H \v_\lambda\,.
\eeq
As one can see, perturbations of all matter species influence the evolution of the Weyl
fluid. In turn, perturbations of the Weyl fluid affect the gravitational potentials via
\eqref{GR Einstein Psi brane multi-component} and \eqref{GR Einstein zeta brane
multi-component}, which influences the perturbations of matter via \eqref{GR conserv rho
brane multi-component} and \eqref{GR conserv upsilon brane multi-component}.

Using \eqref{delta C multicomponent}, \eqref{rho C-1}, \eqref{v C} and \eqref{no-boundary
dim less variable}, we obtain
\beq \label{delta pi C}
\delta \pi_{\C} = -\, \frac{2 \rho_r a^2 (1+3\gamma)}{3 k^2} \left[\delta_{\C} + \frac{6
H \v_\C}{2+\beta} + \frac{3}{4\rho_r}\sum_\lambda (\rho_\lambda +
p_\lambda)\Delta_\lambda \right] \,,
\eeq
\beq\label{v C-2}
\dot \v_\C  = \l(\gamma + \frac{2}{3}\r) \delta_\C + \l(3 \gamma + \frac{4}{2+\beta}\r) H \v_\C + \l(\frac{3\gamma
+ 1}{4\rho_r}\r) \sum_{\lambda}(\rho_\lambda+p_\lambda)\Delta_\lambda  \, .
\eeq

From \eqref{GR Einstein Psi brane multi-component} and \eqref{GR Einstein zeta brane
multi-component}, one can express the gravitational potentials in terms of the variables
$\Delta_\lambda$, $\v_\lambda$, $\delta_\C$ and $\v_\C$:
\ber \label{GR Einstein Psi brane multi-component-2}
\Psi &=& -\,\frac{(2+\beta) a^2}{2 m^2 k^2
\beta}\sum_{\lambda}(\rho_\lambda+p_\lambda)\Delta_\lambda - \frac{4 \rho_r a^2}{3 m^2
k^2 \beta } \left(\delta_\C + 3 H \v_\C \right)\,, \\
\label{GR Einstein zeta brane multi-component-2} \Psi-\Phi &=& -\, \frac{8 \rho_r a^2}{3
m^2 k^2 \beta} \left[\delta_{\C} + \frac{6 H \v_\C}{2+\beta} +
\frac{3}{4\rho_r}\sum_\lambda (\rho_\lambda + p_\lambda)\Delta_\lambda \right]\,.
\eer

The equations for the variables $\Delta_\lambda$ and $\v_\lambda$ can be derived from
\eqref{GR conserv rho brane multi-component} and \eqref{GR conserv upsilon brane
multi-component}:
\ber \label{GR conserv Delta brane multi-component}
\dot{\Delta}_\lambda &=& 3 H c_{s \sss \lambda}^2 \Delta_\lambda -
\frac{k^2}{a^2}\,\v_\lambda + \frac{3(2+\beta)}{2 m^2 \beta} \sum_{\mu \neq
\lambda}(\rho_\mu + p_\mu)(\v_\mu - \v_\lambda) + \frac{4\rho_r}{m^2 \beta}\,\v_\C \,, \\
\label{GR conserv V brane multi-component} \dot{\v}_\lambda  &=& c_{s \sss
\lambda}^2\Delta_\lambda + \frac{(2-\beta)a^2}{2 m^2 k^2 \beta}\sum_\mu (\rho_\mu +
p_\mu)\Delta_\mu + \frac{4 \rho_r a^2}{3 m^2 k^2 \beta} \,\delta_\C + \frac{4 (2-\beta)
\rho_r a^2}{m^2 k^2 \beta (2+\beta)} \,H \v_\C \,.
\eer

Thus, to determine the evolution of cosmological perturbations, we need to solve the
system of equations \eqref{rho C-1}, \eqref{delta C multicomponent}, \eqref{GR conserv
Delta brane multi-component}, \eqref{GR conserv V brane multi-component}. After finding
solutions of this system, one can use \eqref{GR Einstein Psi brane multi-component-2},
\eqref{GR Einstein zeta brane multi-component-2} to determine the gravitational
potentials.

%%%%%%%%%%%%%%%%%%%%%%%%%%%%%%%%%%%%%%%%%%%%%%%%%%%%%%%%%%%%%%%%%%%%%%%%%%%%%%%%%%%%%%%%%%%%%%%%
\subsection{Perturbations of a single fluid }
\label{Sec: Perturbations of a dominant component}

In this subsection, we consider the situation when one fluid component dominates over
the rest. In this case, for the dominating component we can use a
single-fluid version\footnote{Some correction from sub-dominant matter component might be
expected from the term $\propto (\v_\mu-\v_\lambda)$ in \eqref{GR conserv Delta brane
multi-component} in the super-Hubble regime, when spatial gradients are neglected.
However, for the adiabatic modes under consideration in this paper, such corrections are
absent due to appropriate initial conditions; see Eq.~\eqref{brane initial conditions
CDM} below.} of \eqref{GR conserv Delta brane multi-component} and \eqref{GR conserv V
brane multi-component}:\footnote{We omit the label $\lambda$ for the dominating matter
component in this subsection.}
\beq\label{v brane}
\dot \Delta - 3 H \s \Delta = -\, \frac{k^2}{a^2}\,\v + \frac{4 \rho_r \v_\C}{m^2
\beta}\,,
\eeq
\beq \label{GR conserv V brane one-component} \dot{\v}  =
c_{s \sss }^2\Delta + \frac{(2-\beta)a^2}{2 m^2 k^2 \beta}(\rho + p)\Delta + \frac{4
\rho_r a^2}{3 m^2 k^2 \beta} \,\delta_\C + \frac{4 (2-\beta) \rho_r a^2}{m^2 k^2 \beta
(2+\beta)} \,H \v_\C \,.
\eeq

Remarkably, using \eqref{v brane}, \eqref{GR conserv V brane one-component} and \eqref{v
C-2}, one can derive a single second-order differential equation for the variable
$\Delta$:
\ber\label{Delta}
\ddot{\Delta} + (2-3\s) H \dot{\Delta} =
\l[\frac{\rho+p}{2m^2}\l(1+\frac{6\gamma}{\beta}\r) + 3\s\l(\dot{H}+2H^2 -
\frac{k^2}{3a^2}\r) + 3H\dot{(\s)}\r]\Delta \nonumber \\ {} + \frac{4 \rho_r
(1+3\gamma)}{3m^2\beta}\,\,\delta_\C  \,,
\eer
which should be supplemented with \eqref{delta C multicomponent}:
\beq\label{delta C}
\ddot\delta_\C + \l(\frac{2\beta}{2+\beta} - 3\gamma\r)H\dot\delta_\C + \frac{k^2}{3
a^2}(2+3\gamma)\delta_\C = - \,\frac{k^2 (1+3\gamma)}{4\rho_r a^2}(\rho+p)\Delta\,.
\eeq

The system of equations \eqref{Delta}, \eqref{delta C} describes perturbations of
a relativistic fluid with arbitrary equation of state $w=p/\rho$ and
whose adiabatic sound speed $\s$ is given by \eqref{sound speed def}.

Once the solutions for $\Delta$ and $\delta_\C$ are known, one can find the momentum
potentials for matter and Weyl fluid, namely $\v,\v_\C$,
 via the relations \eqref{rho C-1} and \eqref{v
brane}. After that, the gravitational potentials $\Phi$ and $\Psi$ can be determined from
\eqref{GR Einstein Psi brane multi-component-2} and \eqref{GR Einstein zeta brane
multi-component-2}:
\ber \label{GR Einstein Psi brane one component}
\Psi &=& -\,\frac{(2+\beta) a^2}{2 m^2 k^2 \beta}(\rho+p)\Delta - \frac{4 \rho_r a^2}{3
m^2 k^2 \beta } \left(\delta_\C + 3 H \v_\C \right)\,, \\
\label{GR Einstein zeta brane one component} \Psi-\Phi &=& -\, \frac{8 \rho_r a^2}{3 m^2
k^2 \beta} \left[\delta_{\C} + \frac{6 H \v_\C}{2+\beta} + \frac{3}{4\rho_r} (\rho +
p)\Delta \right]\,.
\eer

Finally, the evolution of all other (sub-dominant) fluid components is described by
\eqref{GR conserv rho brane multi-component} and \eqref{GR conserv upsilon brane
multi-component} with the gravitational potentials \eqref{GR Einstein Psi brane one
component}, \eqref{GR Einstein zeta brane one component} as source terms.

%%%%%%%%%%%%%%%%%%%%%%%%%%%%%%%%%%%%%%%%%%%%%%%%%%%%%%%%%%%%%%%%%%%%%%%%%%%%%%%%%%%%%%%%%%%%%%x

\subsection{The Friedmann expansion regime}
\label{sec: general-relativistic regime}

An important feature of our braneworld, which distinguishes it from the Randall--Sundrum
model, is that the effect of the extra dimension on cosmic expansion is usually small at
early times \cite{Shtanov:2000vr, Sahni:2002dx}. We refer to this early epoch as the {\em
Friedmann regime\/}, since the equations of $(3+1)$-dimensional general relativity
determine the course of cosmic expansion during this early time [see
Eq.~(\ref{background-flat})]. Nevertheless, at the perturbative level, perturbations of
the extra-dimensional Weyl fluid exist at all times and can never be
ignored. Thus, we investigate the evolution of perturbations during early times when
\begin{equation}\label{condition_background-gr}
\sum_\lambda\rho_\lambda+\sigma \gg \frac{m^2}{\ell^{2}}\, , \qquad H \gg \ell^{-1}
\, ,
\end{equation} \label{Fr-app}
implying that the effect of the extra dimension (parameterized by the inverse length
$\ell^{-1} = M^3 / 2 m^2$) on background evolution is small.  In this approximation,
equation \eqref{background-flat} turns into the Friedmann expansion law
\beq \label{background-gr}
H^2\approx\frac{1}{3m^2}\l(\sum_\lambda\rho_\lambda + \sigma\r) \, ,
\eeq
so that
\beq \label{happrox}
\dot{H}\approx {} - \frac{\sum_\lambda(\rho_\lambda + p_\lambda)}{2 m^2} \, , \quad
\beta\approx {} - 2 \ell H \, , \quad \gamma\approx
\frac{1}{3}\l(1+\frac{\dot{H}}{H^2}\r)\approx {} - \frac{\sum_\lambda\rho_\lambda (1 +
3w_\lambda) - 2 \sigma}{6\l(\sum_\lambda\rho_\lambda + \sigma\r)} \, .
\eeq

Relation \eqref{delta C}, which describes the evolution of the Weyl-fluid, simplifies to
\beq\label{delta C GR regime}
\ddot\delta_\C + \l(2 - 3\gamma\r)H\dot\delta_\C + \frac{k^2}{3 a^2}(2+3\gamma)\delta_\C
= - \,\frac{k^2(1+3\gamma)}{4\rho_r a^2} \sum_{\lambda}(\rho_\lambda+p_\lambda)\Delta_\lambda\,,
\eeq
where we have neglected terms of order $1 / \beta$ with respect to unity, according
to \eqref{condition_background-gr}, (\ref{happrox}).  Perturbations of the
 energy density in each fluid component under this
approximation can be derived from \eqref{GR conserv Delta brane
multi-component}, \eqref{GR conserv V brane multi-component}:
\ber \label{GR conserv Delta brane multi-component GR regime} \dot{\Delta}_\lambda &=& 3 H c_{s \sss \lambda}^2 \Delta_\lambda - \frac{k^2}{a^2}\,\v_\lambda + \frac{3}{2 m^2} \sum_{\mu\neq\lambda}(\rho_\mu + p_\mu)(\v_\mu - \v_\lambda) + \frac{4\rho_r}{m^2 \beta}\,\v_\C \,,
\\ \label{GR conserv V brane multi-component GR regime} \dot{\v}_\lambda  &=&
c_{s \sss \lambda}^2\Delta_\lambda - \frac{a^2}{2 m^2 k^2}\sum_\mu (\rho_\mu +
p_\mu)\Delta_\mu + \frac{4 \rho_r a^2}{3 m^2 k^2 \beta} \,\l(\delta_\C - 3 \,H \v_\C\r)
\,,
\eer
where the variable $\v_\C$ is related to $\delta_\C$ via \eqref{rho C-1}.

Evolution of the gravitational potentials during the Friedmann regime is determined by
[see \eqref{GR Einstein Psi brane multi-component-2} and \eqref{GR Einstein zeta brane
multi-component-2}]:
\ber \label{GR regime Einstein Psi brane multi-component}
{} - \frac{k^2}{a^2}\,\Psi &=&
\frac{1}{2m^2}\,\sum_{\lambda}(\rho_\lambda+p_\lambda)\Delta_\lambda + \frac{4 \rho_r}{3
m^2 \beta } \left(\delta_\C + 3 H \v_\C \right)\,,
\\ \label{GR regime Einstein zeta brane multi-component}
\Psi-\Phi &=& {} - \frac{2 a^2}{m^2 k^2
\beta}\sum_{\lambda}(\rho_\lambda+p_\lambda)\Delta_\lambda - \frac{8 \rho_r a^2}{3 m^2
k^2 \beta}\l( \delta_\C + \frac{6 H \v_\C}{\beta} \r)\,.
\eer

Finally, we note that in the case of a {\em single-component\/} fluid, we can, instead of
\eqref{GR conserv Delta brane multi-component GR regime} and \eqref{GR conserv V brane
multi-component GR regime}, employ the early-time version of \eqref{Delta}:
\beq\label{Delta GR regime} \ddot \Delta + (2-3\s)
H \dot \Delta  + \l[\dot{H} - 3\s\l(\dot{H}+2H^2 - \frac{k^2}{3a^2}\r) - 3H \dot{\left( \s
\right)} \r]\Delta = \frac{4 \rho_r (1+3\gamma)}{3 m^2 \beta}\,\delta_\C \, .
\eeq

One finds that, in the formal limit of $|\beta| \to \infty$, perturbations of the Weyl
fluid do not affect those of ordinary matter. Thus, we expect that perturbations of
matter components at early times will behave as in general relativity. However, the
approximation $|\beta| \to \infty$ is too crude and does not allow
control of its accuracy [in contrast to the approximation $|\beta|\approx 2\ell H \gg 1$,
which was used to derive \eqref{delta C GR regime}--\eqref{Delta GR regime}].

Below, in Sec.~\ref{sec: Weyl perturbations at the radiation-dominated
epoch}, we analyze the evolution of cosmological perturbations during the epoch of
radiation domination. We shall discover that the approximation (\ref{Delta GR regime}) is
quite accurate during such early times.

%%%%%%%%%%%%%%%%%%%%%%%%%%%%%%%%%%%%%%%%%%%%%%%%%%%%%%%%%%%%%%%%%%%%%%%%%%%%%%%%%%%%%%%%

\subsection{Scaling approximation on super-Hubble scales}
\label{sec: Scaling approximation}

We can observe that, at the early stages of cosmological evolution, when the Friedmann
approximation considered in the previous subsection is applicable, our approach matches
well with the scaling ansatz considered in \cite{Sawicki:2006jj, Song:2007wd,
Seahra:2010fj}. Indeed, in the Friedmann regime, the expansion of the brane is driven  by
the energy density of the dominating matter component [see \eqref{background-gr}] which
evolves by a power law in the scale factor $a$.  One can expect the existence of solution
of \eqref{delta C GR regime} for the variable $\delta_\C$ in the form of a power of $a$
as well. In such a case, we have the order-of-magnitude estimates
\beq\label{SA delta C}
\dot{\delta}_C \sim H \delta_\C \,, \qquad \ddot{\delta}_\C \sim H^2 \delta_\C \,.
\eeq

On super-Hubble scales, where $ k^2 \ll a^2 H^2$, the last term on the left-hand side of
\eqref{delta C GR regime} can be neglected.  If we also neglect the homogeneous part of
solution of \eqref{delta C GR regime} (this condition is equivalent to that there is no
sources for $\delta_\C$ except the brane itself), we obtain
\beq\label{SA delta C Delta}
\delta_\C \sim \frac{k^2(1+3\gamma)\rho}{a^2 H^2\rho_r}\,\Delta \,,
\eeq
where $\rho$ and $\Delta$ are both related to the dominating matter component. To be more
specific, in the era of matter domination, we have $H^2 \propto \rho_m \propto a^{-3}$,
$(1+3\gamma) \approx 1/2 $, and $\Delta_m \propto a$, which results in $\delta_\C
\propto a^3$. In the case of radiation domination, we have $H^2 \propto \rho_r \propto
a^{-4}$, $(1+3\gamma) \propto \rho_m/\rho_r \propto a $ [see also \eqref{background-gr
radiation gamma}], and $\Delta_r \propto a^2$. Correspondingly, $\delta_\C \propto a^5$
in this case.\footnote{ We note that the variable $\delta_\C$ is related to
the master variable (projected onto the brane) $\Theta_b$ via $\Theta_b = -\,\dfrac{4 a^5
\rho_r}{m^2 k^4}\,\delta_\C$ (see \cite{Viznyuk:2013ywa}). As follows from our
consideration, the master variable on the brane behaves as $\Theta_b \propto a^p$, where
$p=6$ in the regime of radiation domination, and $p=4$ if pressureless matter dominates
over radiation. We observe that the powers $p$ in the evolution law of the variable
$\Theta_b$ coincide with those predicted by the scaling ansatz in
\cite{Sawicki:2006jj}.}

Taking into account \eqref{rho C-1}, we also have the order-of-magnitude estimate
\begin{equation}
\v_\C  = {} - \frac{a^2}{k^2}
\, \dot \delta_\C \sim \,\frac{a^2 H}{k^2}
\, \delta_\C  \, . \label{SA V C}
\end{equation}
Then, we can apply \eqref{GR regime Einstein Psi brane multi-component} and \eqref{GR
regime Einstein zeta brane multi-component} to establish the following relation between
the gravitational potentials:
\beq\label{SA Psi Phi}
\frac{\Psi-\Phi}{\Psi} = -\,\frac{2}{\ell H} \l[ 1 + \O\l(\frac{1}{\ell H}\r) + \O\l(\frac{k^2}{a^2 H^2}\r)\r]\,.
\eeq

Thus, in the regime of Friedmann expansion, which is characterized by the condition $\ell
H \gg 1$, perturbations on super-Hubble spatial scales are described by the approximate
relation
\beq\label{SA Psi Phi approx}
\frac{\Psi-\Phi}{\Psi} \approx -\,\frac{2}{\ell H} \,.
\eeq

Relation \eqref{SA Psi Phi approx} can be considered as a super-horizon counterpart of
\eqref{boundary conditions no-boundary}, because, if we {\em assume\/} it, we get a
closed system of equations for perturbations on the brane. Remarkably, the scaling ansatz
for braneworld perturbations gives the same closing relation on super-Hubble scales
\cite{Seahra:2010fj}, which indicates the match between the two methods, at least to the
leading order in a small parameter $1/\ell H \ll 1$.  In this sense, the scaling ansatz,
which is based on the assumption of vanishing bulk master variable on the past Cauchy
horizon, can be regarded as an approximate partial solution of a more general condition
\eqref{boundary conditions no-boundary}.  Our condition \eqref{boundary conditions
no-boundary} allows one to investigate the behavior of perturbations which have been
originated purely in the bulk, along with perturbations originated purely on the brane.
A rigorous definition of these two modes will be given in the next section.

%%%%%%%%%%%%%%%%%%%%%%%%%%%%%%%%%%%%%%%%%%%%%%%%%%%%%%%%%%%%%%%%%%%%%%%%%%%%%%%%%%%%%%%%%%%%%%%%%%%%%%%%%%%%%%%%%%%%%%%%%%%%%%%%%%%%%%%%%%%%%

\section{Perturbations during the radiative epoch}
\label{sec: Weyl perturbations at the radiation-dominated epoch}

\subsection{Initial conditions}
\label{sec: Setting initial conditions}

The primordial spectra for scalar cosmological perturbations are specified deep within
 the
radiation domination epoch. At that time, the modes relevant to structure formation
belong to super-Hubble spatial scales, and perturbations of pressureless
matter\footnote{Pressureless matter in our investigation possesses all characteristics of
cold dark matter.} are decoupled from those of radiation.\footnote{Ultra-relativistic
primordial plasma will be treated as an ideal radiation. Effects related to baryons and
neutrino will not be considered in this work.} As is well known, in general relativity,
adiabatic non-decaying modes\footnote{In this paper, we do not investigate possible
effects of isocurvature modes.} on super-Hubble scales remain almost constant in time, and
are related to the value of the gravitational potential as follows:\footnote{In terms of
energy density contrasts this relation implies: $\frac{\delta\rho_{m(i)}}{\rho_{m(i)}} =
\frac{3}{4} \l( \frac{\delta\rho_{r(i)}}{\rho_{r(i)}} \r) = {} - \frac{3\Phi_{(i)}}{2}$.}
\beq\label{brane initial conditions CDM}
\delta_{m(i)} = \delta_{r(i)} = -\,\frac{3\Phi_{(i)}}{2}\,,\quad \v_{m(i)} =\v_{r(i)} =
\frac{\Phi_{(i)}}{2H}\,,\quad \Psi_{(i)}=\Phi_{(i)}={\rm const}\,,
\eeq
where the subscript `$(i)$' denotes initial values,
 the subscript `$m$' refers to pressureless matter, and the subscript `$r$' to
radiation.

As the effects of the
 extra dimension in our model weaken at early times, we
expect the above relations to also be valid  in our braneworld
during the radiative epoch (this expectation will be confirmed
in the next section).  We also assume that initial linear perturbations of matter and
radiation are random with Gaussian statistics. In this case, they are completely
characterized by the power spectrum $\mathcal{P}_m (k)$, defined as
\beq
\left\langle \delta_{m (i)}({\bf k}) \delta_{m (i)}({\bf k'}) \right\rangle =
\left\langle \delta_{r (i)}({\bf k}) \delta_{r (i)}({\bf k'}) \right\rangle =
\frac{\mathcal{P}_m (k)}{4\pi k^3}\delta(\bf{k}+\bf{k'})\, .
\eeq

Cosmological observations, interpreted within the framework
 of the $\Lambda$CDM model, indicate that
the initial power spectrum is nearly flat. The following parametrization is commonly
used:
\beq\label{Phi power spectrum}
\mathcal{P}_m (k) = A_m \l(\frac{k}{k_{*}}\r)^{n_s-1}\, ,
\eeq
where $k_{*}$ is a pivot scale which, in the Planck data analysis \cite{Ade:2013zuv}, is
chosen to be $k_{*}/a_0 = 0.05\, \mbox{Mpc}^{-1}$, $A_m$ defines the normalization of the
spectrum, and $n_s - 1$ is its slope. The power spectrum is flat if the scalar spectral
index $n_s = 1$, which is very close to the observed value $n_s \approx 0.96$
\cite{Ade:2013zuv}.

We will see in the next section that perturbations $\delta_\C$ of the Weyl fluid, as well
as those of matter, weakly depend on time before Hubble-radius crossing. It is thus
natural to assume that the initial value of the Weyl fluid $\delta_{\C (i)}$ is also
randomly distributed with Gaussian statistics, so that
\beq\label{2-point correlator}
\left\langle \delta_{\C (i)}({\bf k}) \delta_{\C (i)}({\bf k'}) \right\rangle =
\frac{\mathcal{P}_{\C}(k)}{4\pi k^3}\delta(\bf{k}+\bf{k'})\,,
\eeq
where the primordial power spectrum $\mathcal{P}_{\C}(k)$ for the Weyl fluid can, in
principle, be different from $\mathcal{P}_m (k)$ for matter (and
radiation).  Setting aside the issue of generation of primordial perturbations, one can
study the consequences of a general parametrization of the preceding type (\ref{Phi power
spectrum}):
\beq\label{muk power spectrum}
\mathcal{P}_{\C} (k) = A_\C\l(\frac{k}{k_{*}}\r)^{\alpha}\, ,
\eeq
where $k_{*}$ is the same pivot scale as in \eqref{Phi power spectrum}, and $A_{\C}$ is
the normalization of the initial power spectrum for the Weyl fluid.

In this paper we do not discuss possible mechanisms for the
 generation of primordial perturbations in our braneworld,
and therefore cannot tell whether or not the primordial perturbations of the
 Weyl fluid are
correlated with those of matter and radiation.  For simplicity we shall
 assume them to be
statistically independent.  In this case, one can consider the evolution of two basic
modes:
 \begin{description}
   \item[Brane mode] initial perturbations of the Weyl fluid are absent:
\beq\label{brane mode}
A_m = A_0 \,, \qquad A_\C = 0\,;
\eeq
   \item[Bulk mode] initial perturbations in matter \& radiation on the brane are
       absent:
\beq\label{bulk mode}
A_m = 0\,, \qquad A_\C = c A_0\,.
\eeq
 \end{description}
Here, $A_0 = 2.2 \times 10^{-9}$ defines the normalization of the
primordial power spectrum, and $c$ is a numerical constant describing the intensity of
the bulk mode. In the case of superposition of these two modes, the power spectrum of
any quantity will be given by the sum of the corresponding power spectra (since these two
modes are assumed to be statistically independent).

We note that, if the initial adiabatic perturbation for matter/radiation and for
the Weyl
fluid are not assumed to be statistically independent, say, if $\delta_{\C(i)}$ is
proportional to $\delta_{r(i)}$ due to a common mechanism for their primordial
generation (which we do not discuss in this work), then the power spectra would be
calculated by squaring a superposition of the two modes.

In the following section, we consider in more detail the evolution of perturbations on
super-Hubble spatial scales in our braneworld model.

\subsection{Evolution of perturbations prior to Hubble crossing}
\label{sec: Before the HG crossing brane}

At early times, cosmological evolution is dominated by an ultra-relativistic
component (which has equation of state $w_r=c^{2}_{s (r)}=1/3$ and can be treated as
radiation). The background cosmological equation \eqref{brane Hubble cosm param} in this
case can be approximated by the relations:
\beq \label{background-gr no sigma}
H\approx H_0 \sqrt{\Omega_r}\, (1 + z)^2 \,, \qquad \dot{H}\approx {} - 2H^2 \,,\qquad
\gamma\approx {} - \frac{1}{3}\, ,
\eeq
where $\Omega_r$ is the cosmological parameter corresponding to radiation. Radiation
dominates as long as $\rho_r \gg \rho_m$, where $\rho_m$ is the
 pressureless matter density
characterized by the cosmological parameter $\Omega_m$. This condition is valid  as long
as\footnote{Here and below, for numerical estimates we assume the following values of the
cosmological parameters \cite{Sahni:2014ooa}: $\Omega_m = 0.28$, $\Omega_r = 7 \times
10^{-5}$, and $\Omega_\ell = 0.025$. In this case, $\Omega_\sigma \approx 1.036$.}
\beq\label{condition radiation domination}
\frac{a}{a_0} \ll \frac{\Omega_r}{\Omega_m}\simeq 2.5 \times 10^{-4}\,.
\eeq
Obviously, the influence of the cosmological parameter $\Omega_\sigma$ in (\ref{brane
Hubble cosm param}) can also be ignored at this stage of cosmological evolution.

The leading braneworld corrections to \eqref{background-gr no sigma} are of
the order of magnitude $1/\ell H$, where the quantity $\ell H$ is estimated as
\beq\label{ell H radiation dimination}
\ell H \approx \sqrt{\frac{\Omega_r}{\Omega_\ell}}\,\l(\frac{a_0}{a}\r)^2 \gg
\frac{\Omega_m^2}{\sqrt{\Omega_\ell \,\Omega_r^3}} \simeq 8.5\times 10^{5}\,.
\eeq
Since $1/\ell H \ll 1$, the universe expands as in general relativity and
perturbations during the radiative regime can be analyzed using the results of
Sec.~\ref{sec: general-relativistic regime}.

In this case, from \eqref{Delta GR regime} and \eqref{delta C GR regime} we obtain a
system of two equations:
\ber \label{GR Delta radiation brane} \ddot{\Delta}_r + H \dot{\Delta}_r -
\l(2H^2-\frac{k^2}{3a^2}\r)\Delta_r &=& -\, \frac{2 H^2 (1+3\gamma)}{\ell H}\,\delta_\C
\,, \\
\label{weyl radiation-dominated equation} \ddot \delta_\C + 3 H \dot \delta_\C +
\frac{k^2}{3a^2}\,\delta_\C  &=& - \,\frac{k^2 (1+3\gamma)}{3 a^2} \,\Delta_r \, .
\eer
We are going to show that the right-hand sides of these equations can be ignored, by
comparing them to some terms on the left-hand sides of the corresponding equation. To do
this, we need to estimate the factor ($1 + 3\gamma$) more precisely. Taking into account
that the next correction to \eqref{background-gr no sigma} comes from the pressureless
matter component, we compute
\beq \label{background-gr radiation gamma}
1 + 3 \gamma \approx \frac{\Omega_m}{2 \Omega_r} \, \frac{a}{a_0} \ll 1\,.
\eeq

Next, we shall assume that, deep in the radiative regime, the
following condition is satisfied:
\beq\label{condition of homogenity}
(1+3\gamma) \ll \left| \frac{ \delta_\C  }{ \Delta_r } \right| \ll \frac{\ell
H}{(1+3\gamma)}\,.
\eeq
Under this condition, the system of equations \eqref{GR Delta radiation brane},
\eqref{weyl radiation-dominated equation} simplifies to
\ber \label{GR Delta radiation brane before HG crossing}
\ddot{\Delta}_r + H \dot{\Delta}_r - \l(2H^2-\frac{k^2}{3a^2}\r)\Delta_r &\approx& 0 \, ,
\\
\label{weyl radiation-dominated equation before HG crossing} \ddot \delta_\C + 3 H \dot
\delta_\C + \frac{k^2}{3a^2}\,\delta_\C  &\approx& 0 \, .
\eer

Perturbations with any given wavenumber $k$ begin their evolution deep inside
the super-Hubble
regime, where $k \ll a H$. Equations \eqref{GR Delta radiation brane before HG crossing}
and \eqref{weyl radiation-dominated equation before HG crossing} can easily be solved in
the formal limit $k \to 0$:
\beq\label{brane before HG crossing Delta-1}
\Delta_r = \Delta_{r(i)} \l(\frac{a}{a_0}\r)^2 + \Delta_{r(d)} \frac{a_0}{a} \, , \qquad
\delta_\C = \delta_{\C {(i)}} + \delta_{\C {(d)}} \frac{a_0}{a} \, ,
\eeq
where $\Delta_{r(i)}$, $\Delta_{r(d)}$, $\delta_{\C {(i)}}$ and $\delta_{\C {(d)}}$ are
constants of integration.

The terms proportional to $a_0/a$ in \eqref{brane before HG crossing Delta-1} correspond
to decaying modes. The influence of decaying modes on future evolution is assumed
to be negligibly small, so, in what follows, these modes shall be neglected.

Let us perform more careful analysis of \eqref{GR Delta radiation brane before HG
crossing} and \eqref{weyl radiation-dominated equation before HG crossing}, allowing for
non-zero $k$. Introduce a new variable:
\beq\label{s definition}
x\equiv \frac{k}{\sqrt{3}\,a H}=\frac{s}{\sqrt{3\Omega_r}}\l(\frac{a}{a_0}\r)^2\,, \qquad s\equiv
\frac{k}{a_0 H_0} = \frac{2\pi}{\lambda_{0} H_0}\,.
\eeq
Here, $\lambda_{0}$ is the spatial scale of perturbation at the present time. From the
viewpoint of structure formation, the most relevant values of $s$ are
\beq\label{s structure formation}
10^{2} \lesssim s \lesssim 10^{6} \, ,
\eeq
where the lower value corresponds to supercluster scales ($\lambda_0\sim
100$~Mpc), and the upper value corresponds to galactic scales
($\lambda_0 \sim 10$~kpc).

In terms of the variable $x$, equations \eqref{GR Delta radiation brane before HG
crossing} and \eqref{weyl radiation-dominated equation before HG crossing} are written as
\ber \label{GR Delta radiation brane before HG crossing x}
x^2\Delta''_r - \l(2-x^2\r)\Delta_r &\approx& 0 \, ,
\\
\label{weyl radiation-dominated equation before HG crossing x} x^2 \delta''_\C + 2
\delta'_\C + x^2\,\delta_\C  &\approx& 0 \, ,
\eer
where the prime denotes differentiation with respect to $x$. In the region $x\ll 1$
(which corresponds to $k \ll a H$), we can look for solution of \eqref{GR Delta radiation
brane before HG crossing x}, \eqref{weyl radiation-dominated equation before HG crossing
x} in the form of an asymptotic expansion:
\ber\label{brane before HG crossing Delta r x}
\Delta_r &\approx& \delta_{r(i)} x^2 \l[ 1-\frac{x^2}{10} + \mathcal{O}(x^4)\r]\,,\\
\label{brane before HG crossing delta C x} \delta_\C &\approx& \delta_{\C
{(i)}}\l[1-\frac{x^2}{6} + \mathcal{O}(x^4)\r] \,.
\eer
Other variables can be calculated to their leading order as
\beq\label{brane before HG crossing Delta r}
\Delta_r \approx \delta_{r(i)} \l(1-\frac{k^2}{30 a^2 H^2}\r) \frac{k^2}{3 a^2 H^2}\,,
\qquad \delta_r \approx -3H\v_r \approx \delta_{r(i)} \left[ 1 + {\cal O} \left(
\frac{k^2}{a^2 H^2} \right) \right] \, ,
\eeq
\beq\label{brane before HG crossing delta C}
\delta_\C \approx \delta_{\C {(i)}}\l(1-\frac{k^2}{18 a^2 H^2}\r) \approx \delta_{\C
{(i)}} \,, \qquad \v_\C \approx \frac{\delta_{\C {(i)}}}{9 H} \left[ 1 + {\cal O} \left(
\frac{k^2}{a^2 H^2} \right) \right] \,,
\eeq
where we have used \eqref{rho C-1} and \eqref{v brane} and assumed that
\beq\label{delta C condition}
\left| \frac{ \delta_\C }{ \Delta_r  } \right| \ll \frac{3\ell H}{2}\,.
\eeq

Condition \eqref{delta C condition} can be written in the form
\beq\label{delta C initial condition}
\left| \frac{\delta_{\C(i)} }{\delta_{r(i)}} \right| \ll \frac{s^2}{2\sqrt{\Omega_\ell
\Omega_r}} \simeq \left( 3.8 \times 10^2 \right) s^2\,,
\eeq
which seems to be quite a realistic restriction on $\delta_{\C (i)}$ for values of $s$ in
the range \eqref{s structure formation}. Violation of \eqref{delta C initial condition},
in fact, threatens a breakdown of the linear approximation. The validity of \eqref{delta
C initial condition} also ensures condition \eqref{condition of homogenity} at this stage
of cosmological evolution. Thus, the approximate solutions \eqref{brane before HG
crossing Delta r} and \eqref{brane before HG crossing delta C} are justified.

As we see, the variables $\delta_r$ and $\delta_\C$ are both constant in time
for modes which lie outside the Hubble radius during the radiation dominated epoch.
Using \eqref{GR regime Einstein Psi brane multi-component} and
\eqref{GR regime Einstein zeta brane multi-component}, we can now estimate
braneworld effects on the evolution of the gravitational potentials. We start with the
potential $\Psi$:
\beq\label{grav potential Psi before HG crossing brane}
\Psi = {} - \frac{2\delta_{r(i)}}{3} \l( 1 - \frac{4\sqrt{\Omega_r\Omega_\ell}}{s^2}\,
\frac{\delta_{\C {(i)}}}{\delta_{r(i)}} \r) \approx {} - \frac{2\delta_{r(i)}}{3}\,,
\eeq
where the approximation is valid due to \eqref{delta C initial condition}.

To evaluate the potential $\Phi$, we compute
\beq\label{grav potential Phi before HG crossing brane}
\frac{\Psi -  \Phi}{\Psi} \approx  {} - \frac{2}{\ell H} -
\frac{6\sqrt{\Omega_r\Omega_\ell}}{s^2}\, \frac{\delta_{\C {(i)}}}{\delta_{r(i)}} \, .
\eeq
From \eqref{ell H radiation dimination} and \eqref{delta C initial condition}, we find
$|(\Psi - \Phi)/\Psi| \ll 1$ prior to Hubble-radius crossing. Thus neither the
perturbation in radiation nor the gravitational potentials are significantly affected by
the Weyl fluid during this period of cosmological evolution.  As a result, our braneworld
has the same initial relations between the perturbation of radiation and gravitational
potentials as those in general relativity:
\beq\label{brane initial conditions radiation}
\Psi_{(i)} \approx \Phi_{(i)} \approx -\,\frac{2\delta_{r(i)}}{3}\,.
\eeq
Using \eqref{brane before HG crossing Delta r}, we can improve this result to include
corrections in $k/aH$:
\beq\label{GR Phi super-Hubble}
\Psi \approx \Phi \approx -\,\frac{2\delta_{r(i)}}{3} \l( 1-\frac{k^2}{30 a^2 H^2} \r) \,
.
\eeq
In passing, we note that relation \eqref{grav potential Phi before HG
crossing brane} coincides with the scaling ansatz prediction \eqref{SA Psi Phi approx} if
the bulk mode is neglected ($\delta_{\C(i)}=0$).

The evolution of perturbations in pressureless matter, whose background density is
subdominant in the radiative regime, can be determined from the conservation
laws \eqref{GR conserv rho brane multi-component}, \eqref{GR conserv upsilon brane
multi-component}, in which the gravitational potentials $\Psi$ and
$\Phi$ act as sources:
\ber \label{GR conserv CDM rho}
\dot{\delta}_m &=& - \frac{k^2}{a^2}\,\v_m + 3\dot{\Psi} \,, \\
\label{GR conserv CDM upsilon} \dot{\v}_m &=& \Phi \, .
\eer
The system of equations \eqref{GR conserv CDM rho}, \eqref{GR conserv CDM upsilon} can
easily be solved with $\Psi$ and $\Phi$ given by \eqref{GR Phi
super-Hubble}:
\beq\label{Before HG crossing CDM 1}
\delta_m \approx \delta_{m (i)} - \,\frac{7 k^2}{20 a^2 H^2}\,\Phi_{(i)}\,, \qquad \v_m
\approx \frac{\Phi_{(i)}}{2H}\l[1-\,\frac{k^2}{60\, a^2 H^2}\r]\,.
\eeq

For the adiabatic mode, the initial values $\delta_{m (i)}$ and $\delta_{r(i)}$ are equal
[see (\ref{brane initial conditions CDM})]. Thus, in the leading approximation, we have
\beq\label{Before HG crossing CDM}
\delta_m \approx \delta_{m (i)}= {} - \frac{3\Phi_{(i)}}{2}\,, \qquad \v_m \approx
\frac{\Phi_{(i)}}{2H}\,, \qquad \Delta_m \equiv \delta_m + 3 H \v_m \approx {} - \frac{3
k^2}{8 a^2 H^2}\,\Phi_{(i)}\,.
\eeq

The initial power spectra for the statistically independent quantities $\Phi_{(i)}$ and
$\delta_{\C(i)}$ are specified in Sec.~\ref{sec: Setting initial conditions}.

\subsection{Evolution of perturbations after Hubble crossing}
\label{sec: After the HG crossing brane}

After Hubble crossing, when $k > a H$, the behavior of perturbations significantly
changes.  Modes most relevant for structure formation [see \eqref{s structure formation}]
all cross the horizon during the radiation-dominated epoch. In this section, we consider
the evolution of perturbations well after Hubble crossing.

Perturbations during the radiative regime are described by the general system of
equations \eqref{GR Delta radiation brane} and \eqref{weyl radiation-dominated equation}.
First, we obtain the homogeneous solution of these equations (i.e., without their
right-hand sides):
\ber\label{After HG crossing GR brane Delta r}
\Delta^{\rm (hom)}_r &=& 3 \delta_{r(i)} \l[ \frac{\sqrt{3}\,aH}{k} \sin
\l(\frac{k}{\sqrt{3}aH}\r)  - \cos \l( \frac{k}{\sqrt{3}aH} \r) \r] \,,
\\ \label{After HG crossing GR brane delta C}
\delta^{\rm (hom)}_\C &=& \delta_{\C(i)}  \frac{\sqrt{3}\,aH}{k} \sin \l(
\frac{k}{\sqrt{3}\,a H} \r) \,,
\eer
where \eqref{After HG crossing GR brane Delta r} and \eqref{After HG crossing GR
brane delta C} have been matched with the non-decaying solutions \eqref{brane before HG
crossing Delta r} and \eqref{brane before HG crossing delta C} at the moment of
Hubble-radius crossing, in order to fix the constants of integration (keeping in mind that
decaying modes have been neglected). Expression \eqref{After HG crossing GR brane Delta r}
coincides with the general-relativistic solution in this regime.

Let us determine the evolution of modes deep inside the Hubble radius, where $k \gg a H
$. Specifically, we shall restrict ourselves to the time period when
\beq\label{deep in sub-Hubble radiation dominated}
4 \times 10^{3} \ll \frac{a_0}{a} \ll (1.2\times 10^2) s\,,
\eeq
when radiation is still dominating ($\rho_r\gg\rho_m$) and $k \gg a H $. The homogeneous
solutions \eqref{After HG crossing GR brane Delta r} and \eqref{After HG crossing GR
brane delta C} now take the form
\ber\label{deep in sub-Hubble Delta r}
\Delta^{\rm (hom)}_r &=& - 3\, \delta_{r(i)} \cos\l(\frac{k}{\sqrt{3}aH}\r) \,,
\\ \label{deep in sub-Hubble delta C}
\delta^{\rm (hom)}_\C &=& \delta_{\C(i)} \frac{\sqrt{3}\,aH}{k} \sin
\l(\frac{k}{\sqrt{3}\,aH}\r) \,.
\eer

Let us discuss the domain of validity of these solutions.  We note that the right-hand
side of \eqref{GR Delta radiation brane} can be neglected under the
condition\footnote{Here and below, we perform estimates for the ratio of amplitudes of
quantities oscillating around zero. }
\beq\label{initial restriction delta C-1}
\frac{ \delta^{\rm (hom)}_\C }{\Delta^{\rm (hom)}_r } \ll \frac{k^2}{6 a^2
H^2}\l(\frac{\ell H}{1+3\gamma}\r) \quad \Rightarrow \quad \frac{ \delta_{\C(i)} }{
\delta_{r(i)} } \ll \frac{s^3}{\Omega_m \sqrt{3\Omega_\ell}} \simeq 13 s^3\,.
\eeq
Violation of \eqref{initial restriction delta C-1}, again, would threaten the validity of
linear approximation. Thus, the evolution of $\Delta_r$ on sub-Hubble scales is
reasonably described by \eqref{deep in sub-Hubble Delta r}.

The right-hand side of \eqref{weyl radiation-dominated equation} can be neglected under
the condition
\beq\label{deep in sub-Hubble condition of homogenity-1}
\frac{ \delta^{\rm (hom)}_\C }{ \Delta^{\rm (hom)}_r } \gg (1+3\gamma) \quad \Rightarrow
\quad \l(\frac{a_0}{a}\r)^2 \gg \frac{\sqrt{3}\,\Omega_m s}{2\,\Omega_r^{3/2}}\cdot
\frac{\delta_{r(i)}}{\delta_{\C(i)}} \simeq \left(4 \times 10^{5} \right) s\,
\frac{\delta_{r(i)}}{\delta_{\C(i)}} \,.
\eeq
We note that this estimate would always be satisfied on sub-Hubble scales if, initially,
\beq\label{initial restriction delta C-2}
\left| \frac{ \delta_{\C(i)} }{ \delta_{r(i)} } \right| \gtrsim 2.5 \times 10^{-2} s\,.
\eeq
However, estimate \eqref{initial restriction delta C-2} can be violated, which means that
the impact of $\Delta_r$ on the evolution of $\delta_\C$ cannot be neglected. We can
account for this influence by solving \eqref{weyl radiation-dominated equation} with
$\Delta_r$ determined in \eqref{deep in sub-Hubble Delta r}:
\beq \label{weyl radiation-dominated after HG crossing}
\ddot \delta_\C + 3 H \dot \delta_\C + \frac{k^2}{3a^2}\,\delta_\C  =
\frac{\delta_{r(i)}\,k^2 (1+3\gamma)}{a^2} \,\cos\l(\frac{k}{\sqrt{3}\,a\,H}\r) \, .
\eeq

Introducing a new variable $x \equiv k/(\sqrt{3}aH)$ and using relations
\eqref{background-gr no sigma} and \eqref{background-gr radiation gamma}, we can
transform \eqref{weyl radiation-dominated after HG crossing} to
\beq \label{weyl radiation-dominated after HG crossing x}
x^2 \delta''_\C + 2 x \delta'_\C + x^2\delta_\C  = B\,x^3\cos x \, , \qquad B \equiv
\frac{3\sqrt{3}\,\Omega_m \delta_{r(i)}}{2\sqrt{\Omega_r} \,s}\,,
\eeq
where the prime denotes differentiation with respect to $x$. 
One can easily find the general solution of this equation in the limit $x \gg 1$
(corresponding to $k \gg aH$):
\beq\label{deep in sub-Hubble delta C general}
\delta_\C \approx \frac{\delta_{\C(i)}}{x}\,\sin x + \frac{B x^2}{6}\,\sin x =
\delta_\C^{\w} + \delta_\C^{\wf} \,,
\eeq
where we have introduced the notation
\ber\label{delta C Weyl free}
\delta_\C^{\wf} &=& \delta_{r(i)} \l(\frac{k(1+3\gamma)}{2\sqrt{3}\,a H}\r)
\sin\l(\frac{k}{\sqrt{3}\,aH}\r) \,, \\
\label{delta C purely Weyl} \delta_\C^{\w} &=& \delta_{\C(i)}\l(\frac{\sqrt{3}\,aH}{k}\r)
\sin\l(\frac{k}{\sqrt{3}\,aH}\r)  \,.
\eer
Here, the function $\delta_\C^{\wf}$ represents the behavior of $\delta_\C$ in the brane
mode, while $\delta_\C^{\w}$ is its behavior in the bulk mode\footnote{We
should note that expression \eqref{delta C purely Weyl} represents the dominating, but
not full, contribution from the bulk mode. Even if $\delta_{r(i)}=0$, still we have some
contribution to $\delta_C$ from $\Delta_r$, which appears due to the back-reaction from
$\delta_C$ in  \eqref{GR Delta radiation brane}.} [see the definition in Sec.~\ref{sec:
Setting initial conditions}; equations (\ref{brane mode}) and (\ref{bulk mode})]. We
observe that the amplitude of the brane mode grows with time, in contrast to the behavior
of the bulk mode, which has decreasing amplitude in this regime.

As established earlier, the contribution from the bulk mode to the evolution of
$\Delta_r$ is negligibly small, which means that the bulk mode of $\Delta_r$ can be
neglected:
\beq\label{deep in sub-Hubble Delta r general}
\Delta_r \approx \Delta_r^{\wf} = -\, 3\, \delta_{r(i)} \cos\l(\frac{k}{\sqrt{3}aH}\r)
\,.
\eeq

The brane mode of $\delta_\C$ \eqref{delta C Weyl free} also might affect $\Delta_r$ via
\eqref{GR Delta radiation brane}. This effect, however, is negligible because of the
condition
\beq\label{condition of homogenity Delta r after HG}
\frac{ \delta^{\wf}_\C }{ \Delta^{\wf}_r } \ll \frac{k^2}{6 a^2 H^2}\l(\frac{\ell
H}{1+3\gamma}\r) \quad \Rightarrow \quad \frac{\sqrt{3}\,k}{a H} \, \frac{\ell
H}{(1+3\gamma)^2}\gg 1\, ,
\eeq
which is obviously valid in the regime under consideration, characterized by $k \gg a H$,
$\ell H \gg 1$ and $(1+3\gamma)\ll 1$. Consequently, the evolution of $\Delta_r$ is
described by the brane mode \eqref{deep in sub-Hubble Delta r general} with high
accuracy.

Finally, we are in a position to investigate the influence of the Weyl-fluid perturbation
on the gravitational potentials. From \eqref{GR regime Einstein Psi brane
multi-component}, we have
\beq \label{Psi after HG}
{} - \frac{k^2}{a^2}\,\Psi \approx \frac{2\rho_r\,\Delta_r}{3m^2} \l( 1 -
\frac{\delta_\C}{\ell H \Delta_r} \r) \,,
\eeq
where we have used the estimate $H|\v_\C|\ll |\delta_\C|$, which follows from the definition
\eqref{rho C-1}, solution \eqref{deep in sub-Hubble delta C general} and condition $k \gg
a H$. In the regime under consideration, we can also make estimates
\beq \label{Psi after HG condition-1}
\frac{1}{\ell H}  \frac{\delta^{\wf}_\C}{\Delta_r}  =
\frac{\Omega_m\sqrt{\Omega_\ell}\,s}{12\sqrt{3}\,\Omega_r^2}\l(\frac{a}{a_0}\r)^4 \ll 1.7
\times 10^{-9}s \,
\eeq
for the brane mode \eqref{delta C Weyl free}, and
\beq \label{Psi after HG condition-2}
\frac{1}{\ell H} \frac{\delta^{\w}_\C }{\Delta_r }  =
\sqrt{\frac{\Omega_\ell}{3}}\l(\frac{a}{a_0}\r) s^{-1} \l| \frac{\delta_{\C(i)}}{
\delta_{r(i)}} \r|  \ll 2.3 \times 10^{-5} s^{-1} \l| \frac{\delta_{\C(i)}}{
\delta_{r(i)}} \r| \,
\eeq
for the bulk mode \eqref{delta C purely Weyl}.

Consequently, the correction to the general-relativistic result in \eqref{Psi after HG}
is negligibly small if
\beq \label{Psi after HG condition-3}
\left| \frac{\delta_{\C(i)}}{ \delta_{r(i)}} \right| \lesssim 4.3 \times 10^4\,s \qquad
\text{and} \qquad s\lesssim 5.9 \times 10^8\,.
\eeq
Using \eqref{GR regime Einstein zeta brane multi-component}, we can also verify that
\beq \label{Phi after HG}
\left |\frac{\Psi - \Phi}{\Psi} \right | \ll 1
\eeq
under the same conditions \eqref{Psi after HG condition-3}.

For reasonable initial conditions that do not invalidate the linear approximation
[compare with \eqref{delta C initial condition}], and for values of $s$ relevant to
structure formation [see \eqref{s structure formation}], both conditions in \eqref{Psi
after HG condition-3} are satisfied, and we can conclude that the Weyl fluid does not
significantly affect the gravitational potentials during the radiation-dominated epoch.

Perturbations of pressureless matter during radiation domination are described
by the system of equations \eqref{GR conserv CDM rho}, \eqref{GR conserv CDM upsilon}, in
which the gravitational potentials $\Phi$ and $\Psi$ act as sources. Since the
gravitational potentials are unaffected by the Weyl fluid, the same is true for
pressureless matter perturbations, which thus evolve according to the
general-relativistic law \cite{Gorbunov-Rubakov}:
\beq\label{After HG crossing CDM exact}
\delta_m \approx - 9\Phi_{(i)}\l(\log\frac{k}{\sqrt{3}a H} + \texttt{C} - \frac{1}{2}\r)\,,
\eeq
where $\texttt{C}=0.577\ldots$ is Euler's constant.

Summarizing this subsection, we have established the behavior of Weyl-fluid perturbations
in the sub-Hubble regime of the radiation-dominated epoch [relation \eqref{deep in
sub-Hubble delta C general}]. Since the amplitude of the bulk mode in \eqref{deep in
sub-Hubble delta C general} decreases with time, while the amplitude of the brane mode
grows, one expects the brane mode to dominate during future epochs. Matter perturbations
in this regime are dominated by the brane mode, which exhibits general-relativistic
behavior. Deviations from general-relativistic behavior can, however, be caused by the
brane mode during the future matter-dominated epoch, which we discuss in the next
section.

%%%%%%%%%%%%%%%%%%%%%%%%%%%%%%%%%%%%%%%%%%%%%%%%%%%%%%%%%%%%%%%%%%%%%%%%%%%%%%%%%%%%%%%

\section{Perturbations during matter-domination}
\label{sec: Weyl perturbations in the matter-dominated epoch}

After transition from radiation domination to matter domination, we can once more
consider perturbations in a single-component fluid. Thus, to describe the evolution of
cosmological perturbations during this epoch, we use equations \eqref{Delta} and
\eqref{delta C} with $\s=w=0$:
\ber\label{Delta matter domination}
\ddot{\Delta}_m + 2 H \dot{\Delta}_m -
\frac{\rho_m}{2m^2}\l(1+\frac{6\gamma}{\beta}\r)\Delta_m
&=& \frac{4 \rho_r (1+3\gamma)}{3m^2\beta}\,\,\delta_\C \,, \\
\label{delta C matter domination}
\ddot\delta_\C + \l(\frac{2\beta}{2+\beta} - 3\gamma\r)H\dot\delta_\C + \frac{k^2}{3 a^2}(2+3\gamma)\delta_\C
&=& - \,\frac{k^2(1+3\gamma)\rho_m}{4 a^2 \rho_r}\,\Delta_m\,,
\eer
where the time-dependent parameters $\beta$ and $\gamma$ are defined in \eqref{beta-3}
and \eqref{gamma-flat}, respectively.

The system of equations \eqref{Delta matter domination}, \eqref{delta C matter
domination} (in slightly different notation) was thoroughly investigated in
\cite{Viznyuk:2013ywa}. It was argued there that the problem greatly simplifies during the
Friedmann regime of expansion when braneworld effects do not contribute significantly to
cosmic expansion and can be ignored.
This will be discussed in the
next subsection.

%%%%%%%%%%%%%%%%%%%%%%%%%%%%%%%%%%%%%%%%%%%%%%%%%%%%%%%%%%%%%%%%%%%%%%%%%%%%%%%%%%%%%%%%%%%

\subsection{Matter perturbations during the Friedmann regime}
\label{sec: Matter-dominated epoch general-relativistic regime}

Background cosmological evolution during the matter-dominated Friedmann regime
is described by the approximate relations of Sec.~\ref{sec: general-relativistic
regime}:
\beq \label{background-gr no sigma matter}
H\approx H_0 \sqrt{\Omega_m}\, (1 + z)^{3/2} \,, \qquad \dot{H}\approx {} - \frac{3}{2}\,H^2 \,,\qquad
\gamma\approx {} - \frac{1}{6}\, ,
\eeq
where $\Omega_m$ is the cosmological parameter corresponding to pressureless matter. One
should note that the effect of the cosmological constant has been neglected, which
is valid while
\beq\label{lambda term neglected}
(1 + z)^{3}\gg \frac{\Omega_\sigma}{\Omega_\m} \approx 3.7\,.
\eeq
Using \eqref{background-gr no sigma matter}, we can estimate:
\beq\label{ell H matter domination}
\ell H\approx \sqrt{\frac{\Omega_m}{\Omega_\ell}}\, (1 + z)^{3/2}\approx 3.3\, (1 +
z)^{3/2}\,.
\eeq
Obviously, the validity of \eqref{lambda term neglected} justifies the
early-time Friedmann regime [see condition \eqref{condition_background-gr}] and,
therefore, relations \eqref{background-gr no sigma matter}.

The early-time version of
\eqref{Delta matter domination}, \eqref{delta C matter domination}, which is valid when $z \gg 1$,
reads:
\ber\label{Delta matter domination g-r}
\ddot{\Delta}_m + 2 H \dot{\Delta}_m -
\frac{\rho_m}{2m^2}\,\Delta_m
&=& -\,\frac{\rho_r}{3m^2 \ell H}\,\,\delta_\C \,, \\
\label{delta C matter domination g-r}
\ddot\delta_\C + \frac{5 H}{2}\,\dot\delta_\C + \frac{k^2}{2 a^2}\,\delta_\C
&=& - \,\frac{k^2\rho_m}{8 a^2 \rho_r}\,\Delta_m\,.
\eer

The system of equations \eqref{Delta matter domination g-r}, \eqref{delta C matter
domination g-r} is valid on all spatial scales.  We consider small and large spatial
scales separately.

\subsubsection{Matter perturbations on super-Hubble spatial scales}
\label{sec: Matter-domination super-Hubble}

The behavior of perturbations on super-Hubble spatial scales in matter-dominated regime
is important for the large-scale modes that remain outside the horizon during the whole
radiation-dominated epoch.

For the zeroth order solution, by setting $k = 0$ and neglecting the decaying mode, we
have the following approximate solution of \eqref{delta C matter domination g-r}:
\beq\label{mstter SH delta C}
\delta_\C \approx \delta_{\C(i)} \,.
\eeq
The constant $\delta_{\C(i)}$ in this approximation is the same as in \eqref{brane before
HG crossing Delta-1}.

Taking into account \eqref{background-gr no sigma matter} and \eqref{mstter SH delta C},
we can present the general solution of \eqref{Delta matter domination g-r} in the form
\beq\label{mstter SH Delta m}
\Delta_m \approx {\cal M}_{0}\,a + \frac{\rho_r}{\ell H\rho_m}\, \delta_{\C(i)} \,,
\eeq
where the decaying mode is again neglected. The first term here ($\propto a$) represents
the usual general-relativistic result, and ${\cal M}_{0}$ is a constant of integration
related to the initial value of matter perturbation $\delta_{r(i)}$ from \eqref{brane
before HG crossing Delta r}. The second term (which evolves as $\sqrt{a}$) is a
correction from the bulk mode (the definitions of the brane and bulk modes were given in
Sec.~\ref{sec: Setting initial conditions}).

At the second step of iteration, we can substitute \eqref{mstter SH Delta m} into the
right-hand side of \eqref{delta C matter domination g-r} to find a correction to the
zero-order result \eqref{mstter SH delta C}. In doing so, we are mostly interested in the
correction coming from the component $\Delta_m = {\cal M}_{0}\,a$, because it gives a
leading contribution from the brane mode to the perturbation of the Weyl fluid. As a
result, we obtain
\beq\label{mstter SH delta C brne}
\delta_\C \approx \delta_{\C(i)} - \,\frac{k^2\rho_m}{96 \,a^2H^2\rho_r}\,{\cal M}_{0}\,a \,.
\eeq

We note that the brane mode of the Weyl fluid (corresponding to $\delta_{\C(i)}=0$)
behaves in accordance with the scaling ansatz considered in \cite{Sawicki:2006jj,
Song:2007wd, Seahra:2010fj} and discussed in Sec.~\ref{sec: Scaling approximation}:
$\delta_{\C} \propto a^3$. Solutions \eqref{mstter SH Delta m} and \eqref{mstter SH delta
C brne} can be used to evaluate the gravitational potentials via \eqref{GR regime
Einstein Psi brane multi-component} and \eqref{GR regime Einstein zeta brane
multi-component}:
\beq\label{SA Psi Phi approx matter domination}
\frac{\Psi-\Phi}{\Psi} \approx -\,\frac{2}{\ell H} \,,
\eeq
which confirms our previous estimate \eqref{SA Psi Phi approx}.

We observe that deviations of the evolution of brane perturbations from the
general-relativistic behavior is insignificant on super-Hubble spatial scales (due to the
relation $\ell H \gg 1$ valid in the Friedmann regime under consideration) {\em provided
only the brane mode is taken into account\/}. The possible presence of the bulk mode
requires special attention.  In the following section, we will study the effect of the
bulk mode for perturbations already in the sub-Hubble regime at the moment of transition
from radiation domination to matter domination.

\subsubsection{Matter perturbations on sub-Hubble spatial scales}
\label{sec: Matter-domination sub-Hubble}

It was shown in \cite{Viznyuk:2013ywa} that, deep in the matter-dominated regime, and for
sufficiently large amplitudes of $\delta_\C$, the right-hand side of \eqref{delta C
matter domination g-r} can be neglected both on super-Hubble and sub-Hubble spatial
scales,\footnote{ This is not true at later times of cosmological evolution,
when the quasi-static approximation is applicable. For the quasi-static approximation,
see Sec.~\ref{sec: Koyama-Maartens approximation}.} and only the homogeneous part of the
general solution for the Weyl-fluid perturbations in this regime can significantly
influence the dynamics of matter perturbations. On sub-Hubble spatial scales, the
solution can be presented in the form
\beq\label{delta C matter domination g-r solution}
\delta_\C \approx \l(\frac{a H}{k}\r)^{3/2} \l( F \cos \frac{\sqrt{2}k}{a H} + G \sin
\frac{\sqrt{2}k}{a H} \r)\,,
\eeq
where the constants of integration $F$ and $G$ can be calculated via matching
\eqref{delta C matter domination g-r solution} with \eqref{delta C purely Weyl} for the
bulk mode and with \eqref{delta C Weyl free} for the brane mode at the matter--radiation
equality.

Using \eqref{delta C matter domination g-r solution}, we can solve \eqref{Delta matter
domination g-r} and find the correction from the bulk mode to the general-relativistic
result $\Delta_m = {\cal M} a$ on sub-Hubble spatial scales:
\beq \label{Delta matter domination g-r solution}
\Delta_m \approx {\cal M} a + \frac{2\Omega_r\sqrt{\Omega_\ell}}{\Omega_m \,s} \l(\frac{a
H}{k}\r)^{5/2} \l( F \cos \frac{\sqrt{2}k}{a H} + G \sin \frac{\sqrt{2}k}{a H} \r) \, ,
\eeq
where $s$ was defined in \eqref{s definition}.

We note here that both the brane mode and bulk mode give contribution to the constants
${\cal M}$, $F$, and $G$ in \eqref{Delta matter domination g-r solution}.
(See (\ref{brane mode}), (\ref{bulk mode}) for a description of brane and bulk modes.)
However, as
will be revealed in the following analysis, the contribution of the bulk mode to the
constant ${\cal M}$ is much smaller than that of the brane mode for all reasonable
initial conditions. The Weyl-fluid perturbation \eqref{delta C matter domination g-r
solution} decreases as $a^{-3/4}$, and the oscillatory perturbation of pressureless
matter [the second term in \eqref{Delta matter domination g-r solution}] as $a^{-5/4}$.
The brane mode $\Delta^{\wf}_m = {\cal M} a$ in \eqref{Delta matter domination g-r
solution} is thus dominating in the full solution for pressureless matter perturbations
at early times. Thus, significant deviation from the
early-time regime, (\ref{delta C matter domination g-r solution}), can be expected only
at late times, when the condition $H \gg \ell^{-1}$ ceases to be valid and
bulk effects come into play. Equations \eqref{Delta matter domination} and \eqref{delta
C matter domination} at this period of evolution can be integrated only numerically.

More accurate results for perturbation growth can be obtained by numerically integrating
the exact system of equations, including perturbations of matter, radiation and the Weyl
fluid, starting from the time well before Hubble-radius crossing. This procedure will be
implemented in Sec.~\ref{sec: Numerical integration}\@.

%%%%%%%%%%%%%%%%%%%%%%%%%%%%%%%%%%%%%%%%%%%%%%%%%%%%%%%%%%%%%%%%%%%%%%%%%%%%%%%%%%%%%%%%%%

\subsection{The Quasi-static approximation}
\label{sec: Koyama-Maartens approximation}

In the quasi-static approximation, time derivatives of perturbations in the Weyl fluid
are assumed to be much smaller than spatial gradients (on sub-Hubble scales) and are
neglected [this reduces \eqref{boundary conditions no-boundary} to \eqref{boundary
conditions qs}]. Under this condition, Eq.~\eqref{delta C matter domination} reduces to
\beq\label{quasi-static delta C}
\delta_\C \approx {} - \frac{3 \rho_m (1+3\gamma)}{4 \rho_r (2+3\gamma)}\,\Delta_m\,,
\eeq
which results in a closed equation for $\Delta_m$ in \eqref{Delta matter domination}, namely
\beq\label{quasi-static Delta m}
\ddot{\Delta}_m + 2 H \dot{\Delta}_m \approx
\frac{\rho_m \Delta_m}{2m^2}\l(1+\frac{1}{3 \mu}\r)
 \,.
\eeq
Here, we have introduced a new parameter $\mu$ to put this equation in the form in which
it is presented in \cite{Koyama:2005kd}:
\beq\label{quasi-static mu definition}
\mu \equiv 1+\ell H \l(1+\frac{\dot{H}}{3H^2}\r) = {} - \frac{\beta(2+3\gamma)}{6}
 \,.
\eeq

The evolution of the gravitational potentials in the quasi-static approximation can be
determined from \eqref{GR Einstein Psi brane one component} and \eqref{GR Einstein zeta
brane one component}:
\ber\label{quasi-static Psi}
{} - \frac{k^2}{a^2}\, \Psi \ &\approx& \ \frac{\rho_m \Delta_m}{2 m^2} \l(1 - \frac{1}{3\mu} \r)\,, \\
\label{quasi-static Phi} {} - \frac{k^2}{a^2}\, \Phi \ &\approx& \ \frac{\rho_m
\Delta_m}{2 m^2} \l( 1 + \frac{1}{3\mu} \r)\,.
\eer

Note that, in the quasi-static approximation, one obtains a closed second-order
differential equation \eqref{quasi-static Delta m} for matter perturbation which does not
depend on the spatial scale explicitly [see \eqref{Delta matter domination} and
\eqref{delta C matter domination} for comparison], although the initial conditions can be
scale-dependent. From \eqref{quasi-static Psi} and \eqref{quasi-static Phi} it is evident
that, under the quasi-static approximation, $(\Psi-\Phi)/\Psi$ and $\Phi/\Psi$ are also
scale independent,
\begin{equation}\label{eq:Psi_Phi_frac}
\frac{\Psi-\Phi}{\Psi}=-\frac{2}{3\mu-1}\, ,\qquad \frac{\Phi}{\Psi}=\frac{3\mu+1}{3\mu -1}\, .
\end{equation}

The quasi-static approximation implies that only the brane mode is significant for matter
perturbations at late times, while the contribution from the bulk mode appears to be
negligibly small. This result is consistent with the considerations of Secs.~\ref{sec:
After the HG crossing brane} and \ref{sec: Matter-dominated epoch general-relativistic
regime}, where we have seen that the bulk mode rapidly decays after the Hubble-radius
crossing. Numerical integration of the exact system of equations also confirms the
convergence to the quasi-static approximation, as will be demonstrated in Sec.~\ref{sec:
Numerical integration}.  This is also in agreement with the results of
\cite{Sawicki:2006jj, Song:2007wd, Cardoso:2007xc, Seahra:2010fj}, which confirm the
quasi-static regime on small spatial scales.

%%%%%%%%%%%%%%%%%%%%%%%%%%%%%%%%%%%%%%%%%%%%%%%%%%%%%%%%%%%%%%%%%%%%%%%%%%%%%%%%%%%%%%%%%%%%%%

\section{Results of numerical integration}
\label{sec: Numerical integration}

In Sec.~\ref{Sec: Multi-component fluid on the brane}, we derived a closed system of
equation for perturbations of a multi-component fluid on the brane. These equations can
be numerically integrated, and one can obtain the joint evolution of perturbations of
pressureless matter, radiation and the Weyl fluid.\footnote{In this paper, we
treat the pressureless matter as a single component, neglecting all effects specific to
baryonic matter.}

Equations describing perturbations of matter and radiation densities can be derived from
\eqref{GR conserv Delta brane multi-component} and \eqref{GR conserv V brane
multi-component}:
\beq \label{m-r parity GR regime Delta matter}
\dot{\Delta}_m = {} - \frac{k^2}{a^2}\,\v_m + \frac{2(2+\beta)\rho_r}{m^2 \beta} \,(\v_r
- \v_m) + \frac{4\rho_r}{m^2 \beta}\,\v_\C \,,
\eeq
\beq \label{m-r parity GR regime V matter}
\dot{\v}_m  = \frac{(2-\beta)a^2}{2 m^2 k^2 \beta} \l(\rho_m\Delta_m +
\frac{4\rho_r}{3}\Delta_r \r) + \frac{4 \rho_r a^2}{3 m^2 k^2 \beta} \,\delta_\C  +
\frac{4 (2-\beta) \rho_r a^2}{m^2 k^2 \beta (2+\beta)} \,H \v_\C \,,
\eeq
\beq \label{m-r parity GR regime Delta radiation}
\dot{\Delta}_r = H \Delta_r - \frac{k^2}{a^2}\,\v_r + \frac{3(2+\beta)\rho_m}{2 m^2
\beta} \,(\v_m - \v_r) + \frac{4\rho_r}{m^2 \beta}\,\v_\C \,,
\eeq
\beq \label{m-r parity GR regime V radiation}
\dot{\v}_r  = \frac{\Delta_r}{3} + \frac{(2-\beta)a^2}{2 m^2 k^2 \beta} \l(\rho_m\Delta_m
+ \frac{4\rho_r}{3}\Delta_r \r) + \frac{4 \rho_r a^2}{3 m^2 k^2 \beta} \,\delta_\C  +
\frac{4 (2-\beta) \rho_r a^2}{m^2 k^2 \beta (2+\beta)} \,H \v_\C \,,
\eeq
while the variables $\delta_\C$ and $\v_\C$ evolve according to \eqref{delta C
multicomponent} and \eqref{rho C-1}:
\beq\label{delta C GR regime m-r}
\ddot\delta_\C + \l(\frac{2\beta}{2+\beta} - 3\gamma\r)H\dot\delta_\C + \frac{k^2}{3
a^2}(2+3\gamma)\delta_\C = {} - \frac{k^2(1+3\gamma)}{4\rho_r a^2} \l(\rho_m\Delta_m +
\frac{4\rho_r}{3}\Delta_r \r)\,,
\eeq
\beq\label{v C via rho C m-r}
{} - \frac{k^2}{a^2}\,\v_\C = \dot \delta_\C \,.
\eeq

The parameters $H$, $\beta$ and $\gamma$ are determined by the background cosmological
equations \eqref{brane Hubble cosm param}--\eqref{gamma-flat}, in which the pressureless
matter density $\rho_m$ and the radiation density $\rho_r$ are expressed in terms of
$\Omega_m$ and $\Omega_r$, respectively.

The system of equations \eqref{m-r parity GR regime Delta matter}--\eqref{v C via rho C
m-r} can be numerically integrated from an early epoch when
the relevant mode was in the super-Hubble regime $ k \ll aH$. The initial behavior of
mode functions during this period was derived in Sec.~\ref{sec:
Before the HG crossing brane} [see \eqref{brane before HG crossing Delta r}, \eqref{brane
before HG crossing delta C}, \eqref{brane initial conditions radiation} and \eqref{Before
HG crossing CDM}], and is specified by two initial amplitudes $\delta_{m(i)} =
\delta_{r(i)}$, $\delta_{\C (i)}$ with power spectra \eqref{Phi power spectrum} and
\eqref{muk power spectrum}, respectively.

After the integration of equations \eqref{m-r parity GR regime Delta matter}--\eqref{v C via
rho C m-r}, one determines the gravitational potentials by using the general relations
\eqref{GR Einstein Psi brane multi-component-2} and \eqref{GR Einstein zeta brane
multi-component-2}, which become, respectively:
\ber \label{Psi brane m-r}
\Psi &=& {} - \frac{(2+\beta) a^2}{2 m^2 k^2 \beta}\Bigl( \rho_m\Delta_m +
\frac{4\rho_r}{3}\,\Delta_r \Bigr) - \frac{4 \rho_r a^2}{3 m^2 k^2 \beta }
\Bigl(\delta_\C + 3 H \v_\C \Bigr)\,, \\
\label{zeta brane m-r} \Psi-\Phi &=& {} - \frac{8 \rho_r a^2}{3 m^2 k^2 \beta}
\left(\Delta_r + \frac{3\rho_m}{4\rho_r}\,\Delta_m +  \delta_{\C} + \frac{6 H
\v_\C}{2+\beta} \right)\,.
\eer

\begin{figure}[htb]
\centering
\includegraphics[width=0.55\textwidth]{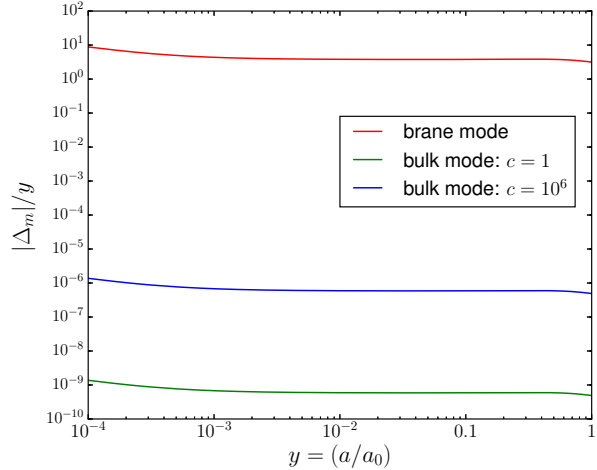}
\caption{Density perturbations on the brane are shown for three cases: (i) Topmost (red)
line shows the evolution of matter density perturbations when perturbations in the
Weyl fluid are initially absent (the brane mode). (ii) The middle (blue) line shows
matter density perturbations when initial perturbations in matter are absent but
perturbations in the Weyl fluid are present (the bulk mode). (iii) The lower (green) curve is the same as
(ii) but with a suppressed initial amplitude for perturbations in the Weyl fluid. In all
three cases, $\Delta_m \propto a(t)$ during the matter-dominated epoch.  Here, the constant $c$
is defined in \eqref{bulk mode}.  One can see that, even for extremely
large initial perturbations of the Weyl fluid (with $c = 10^6$), the brane mode dominates
during the late-time evolution.  For numerical illustration, we have chosen $\Omega_{\ell}=0.025$
and $s=2\pi \cdot 400$ which corresponds to $\sim 10$~Mpc scale.
[See (\ref{brane mode}) and (\ref{bulk mode}) for a description of brane and bulk modes.]
} \label{fig: matter-weyl}
\end{figure}

\begin{figure}[htb]
\centering
\includegraphics[width=0.55\textwidth]{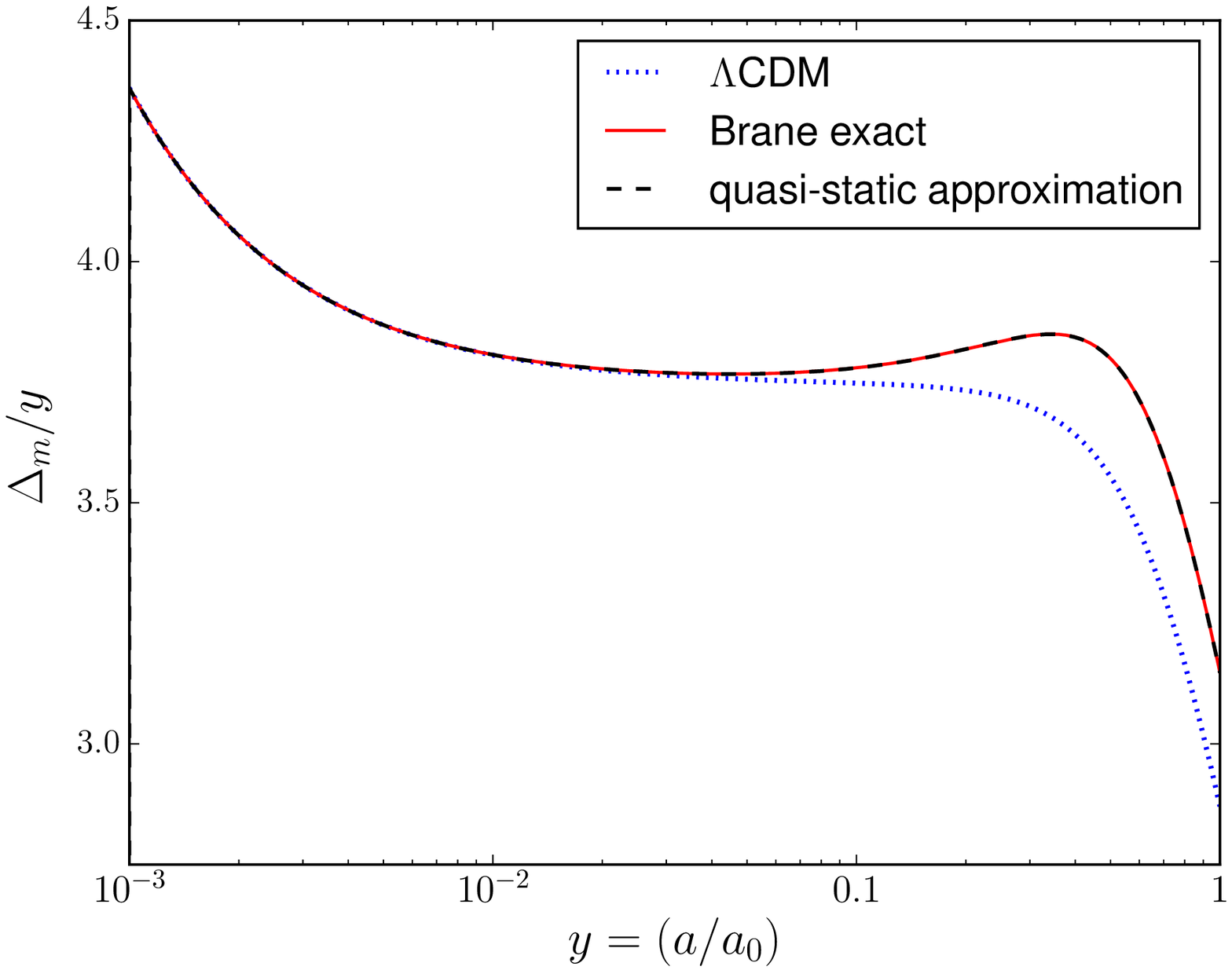}
\caption{Perturbations on the brane obtained using
the exact system of equations (solid red) are compared with those in $\Lambda$CDM (dotted blue)
and the quasi-static
approximation of Koyama--Maartens (dashed black).
The brane parameter is $\Omega_{\ell}=0.025$ and $s \equiv k/a_0H_0 =2\pi \cdot 400 \simeq 2500$ is
the comoving wavelength. Also see figures \ref{fig:scale} and \ref{fig:scale1}, in which the
scale dependence of brane perturbations is highlighted.
}
\label{fig: matter_KM}
\end{figure}

In figure~\ref{fig: matter-weyl}, we present the results of our numerical integration for
late-time perturbations of pressureless matter for the brane mode and for the bulk modes
with two amplitudes of the power spectrum. For simplicity, we choose the scale-invariant
initial power spectrum for the Weyl fluid perturbation, i.e., we set $\alpha=0$ in
\eqref{muk power spectrum}, while numerically calculating the bulk mode. One can see
that, even for an extremely high initial perturbations of $\delta_\C$, the late-time
perturbations of matter are dominated by the brane mode. As long as the first condition
in \eqref{Psi after HG condition-3} is satisfied, the bulk-mode contribution will be
insignificant compared to that of the brane mode. [See (\ref{brane mode}) and (\ref{bulk
mode}) for a description of brane and bulk modes.]

\begin{figure}[htb]
\centering
\includegraphics[width=.55\textwidth]{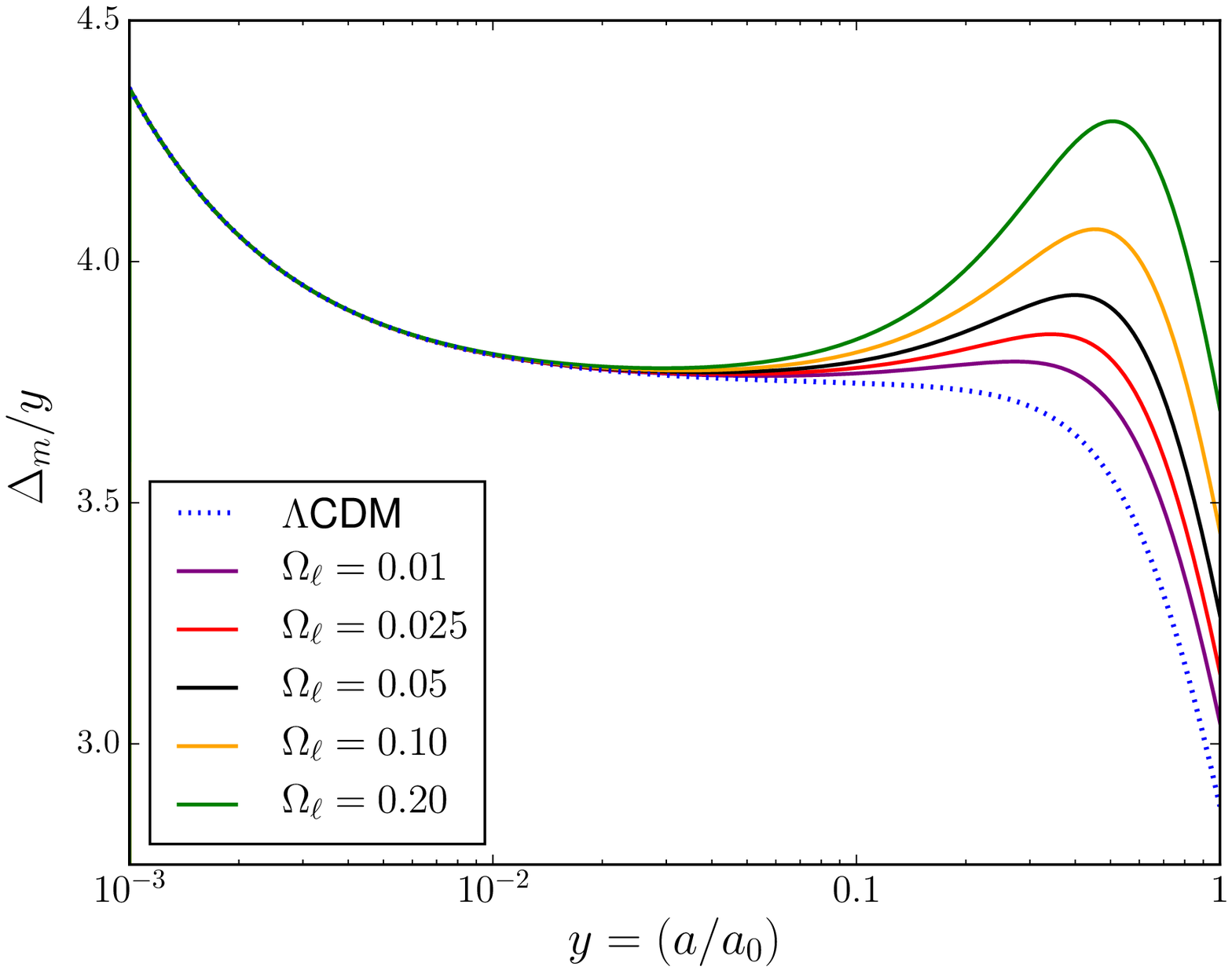}
\caption{Density perturbations on the brane are very sensitive to the value of the brane
parameter $\Omega_\ell$, which depends on the ratio of the five- and four-dimensional
gravitational couplings, see (\ref{beta}) and (\ref{eq:cosmo_parameters}). At late times,
perturbations on the brane grow more rapidly than in $\Lambda$CDM (dotted blue line). The
above results correspond to the scale $s \equiv k/a_0H_0 \simeq 2500$. Unlike
perturbations in $\Lambda$CDM, perturbations on the brane are very weakly scale
dependent. This is illustrated in figure \ref{fig:scale}, in which $\Omega_{\ell}$ is
held fixed while $s$ is allowed to vary, and in figure \ref{fig:scale1}, in which both
$s$ and $\Omega_{\ell}$ are varied. } \label{fig:ol-dep}
\end{figure}

\begin{figure}[htb]
\centering
\includegraphics[width=.55\textwidth]{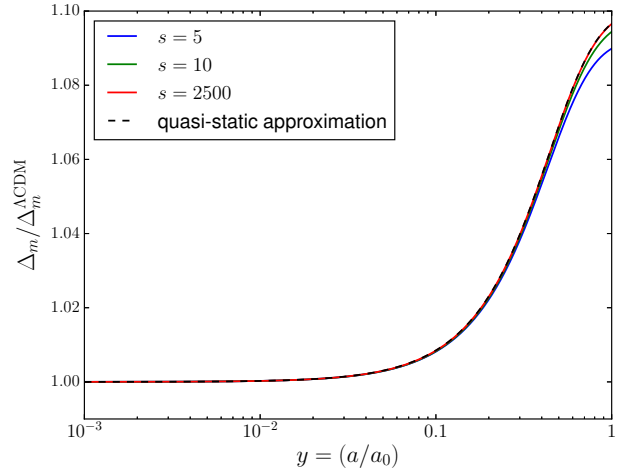}
\caption{Density perturbations on the brane are shown relative to those in $\Lambda$CDM.
Perturbations on the brane are weakly scale-dependent.  The parameter $s
= k / a_0 H_0$ is the ratio of the Hubble length scale to the comoving length scale. We
set $\Omega_{\ell}=0.025$ for all values of $s$.
Note that the evolution on larger spatial scales, with $s \gtrsim 100$, saturates and
agrees with the quasi-static approximation extremely well.
} \label{fig:scale}
\end{figure}

\begin{figure}[htb]
\centering
\includegraphics[width=.55\textwidth]{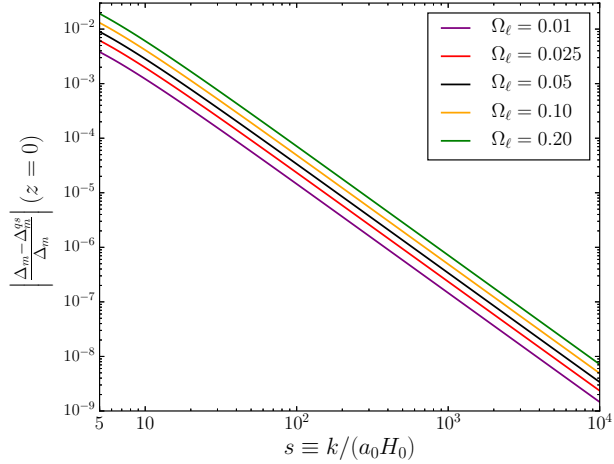}
\caption{The fractional difference between perturbations on the brane
obtained by solving the exact system of equations ($\Delta_m$) and by using the
quasi-static approximation ($\Delta_m^{\rm qs}$) is shown at the present epoch ($z=0$) for
different values of $\Omega_{\ell}=0.025$ and $s = k / a_0 H_0$. One notes that the
accuracy of the quasi-static approximation increases for higher values of $s$ and lower
values of $\Omega_{\ell}$. (The limit $\Omega_{\ell} \to 0$ corresponds to $\Lambda$CDM.)
} \label{fig:scale1}
\end{figure}

\begin{figure}[!htb]
\centering
\includegraphics[width=.55\textwidth]{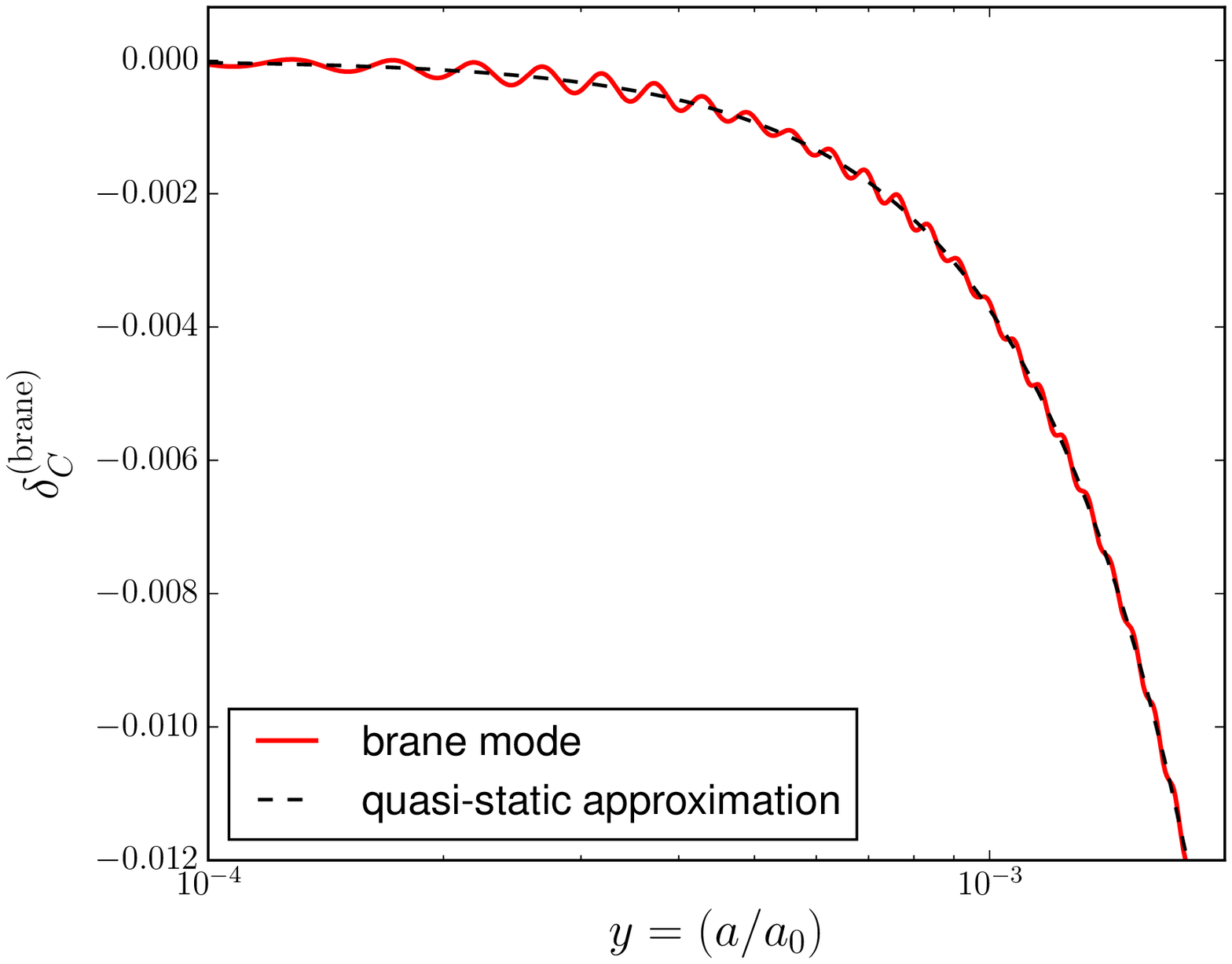}
\caption{Perturbations of the Weyl fluid (in the brane mode) for $\Omega_{\ell}=0.025$ and
$s \equiv k/a_0H_0 \simeq 2500$. The slow growth of the Weyl-fluid perturbations ensures
the validity of the quasi-static approximation. Small oscillations at early times are caused
by the back-reaction from perturbations in radiation, described by the $\Delta_r$ term in
\eqref{delta C Weyl free}. Note that they are not captured by the quasi-static approximation.
} \label{fig:weyl_KM_zoom}
\end{figure}

Concentrating then on the brane mode, we find that late-time perturbations of matter
follow quite well the quasi-static approximation due to Koyama and Maartens, described in
Sec.~\ref{sec: Koyama-Maartens approximation} by equation \eqref{quasi-static Delta m}.
This behavior is illustrated  in figure~\ref{fig: matter_KM} for the wave-number $s=2\pi
\cdot 400 \simeq 2500$, which corresponds to the spatial scale of 10~Mpc. This figure
shows: (i) the growth of perturbations on the brane can differ significantly from those
in $\Lambda$CDM. (ii) Perturbations on the brane with wavelength of 10~Mpc match the
quasi-static approximation quite well. As one increases $\Omega_{\ell}$, perturbations on
the brane show greater departure from those in $\Lambda$CDM. This is illustrated in
figure~\ref{fig:ol-dep} for the single spatial scale of 10~Mpc. This dependency on
$\Omega_\ell$ provides a potent observational test for the braneworld. (Note that in the
limit $\Omega_\ell \rightarrow 0$, the braneworld reduces to $\Lambda$CDM). The
difference between the braneworld and $\Lambda$CDM is shown in figure~\ref{fig:scale} for
various scales. One finds that the ratio $\Delta_m/\Delta_m^{\Lambda CDM}$ increases very
 slowly for smaller scales (higher values of $s$) and
saturates for $s \gtrsim 100$, which includes the range of scales of interest, given in
\eqref{s structure formation}. This indicates the self-similarity of $\Delta_m$ for these
scales. Note that under the quasi-static approximation, the ratio
$\Delta_m/\Delta_m^{\Lambda CDM}$ does not depend upon spatial scale, as
Eq.~\eqref{quasi-static Delta m} differs from the corresponding equation of $\Lambda$CDM
only by a scale-independent factor $[1+1/(3\mu)]$. Hence, in view of
figure~\ref{fig:scale}, one can conclude that the quasi-static approximation is most
accurate on smaller spatial scales while being less accurate on very large spatial scales
($s < 100$). This conclusion is explicitly evident from figure
\ref{fig:scale1}, where the accuracy of the quasi-static approximation is
shown\footnote{In figure \ref{fig:scale1}, we calculate $\Delta_m$ and
$\Delta_m ^{\rm qs}$ with the same initial conditions, $\Delta_m=a/a_0$ and $\dot \Delta_m=\dot a/a_0$,
while starting from $a/a_0=0.001$. Furthermore, we calculate $\Delta_m$ using the system of
equations \eqref{Delta matter domination} and \eqref{delta C matter domination}
valid during matter domination. %The reason behind this is the following: we cannot
%compare the exact solution (using \eqref{m-r parity GR regime Delta matter}--\eqref{v C
%via rho C m-r}) with quasi-static approximation for various $s$ with the initial
%conditions given inside radiation domination.
} as a function of the dimensionless wavenumber $s\equiv k/(a_0 H_0)$ for various values
of $\Omega_{\ell}$. This is the manifestation of the fact that, on smaller spatial
scales (higher $k$ or $s$), the derivatives of $\delta_\C$ are more strongly suppressed
in \eqref{delta C matter domination}.

From \eqref{delta C purely Weyl} and \eqref{delta C matter domination g-r solution} it is
clear that the Weyl fluid perturbation, $\delta_\C$, in the bulk mode, will oscillate
with decaying amplitude inside the horizon, and that this will remain true for any
initial $\delta_{\C(i)}$. During the matter domination epoch, $\delta_\C$ (in both brane
and bulk modes) grows due to the back reaction from $\Delta_m$ in \eqref{delta C matter
domination}, but this effect is expected to be significant only at late times when
$\Delta_m$ has grown sufficiently. However the growth in $\delta_\C$ is slow enough to
satisfy the quasi-static approximation \eqref{quasi-static delta C}, as shown in
figure~\ref{fig:weyl_KM_zoom} for the brane mode. Note that the small oscillations in
$\delta_\C$, shown in figure~\ref{fig:weyl_KM_zoom}, are triggered by the back reaction
from sub-horizon scale oscillating perturbations in radiation, $\Delta_r$,
 during the radiative epoch, see \eqref{delta C
Weyl free}.

\begin{figure}[htb]
\centering
\subfigure[]{
\includegraphics[width=0.488\textwidth]{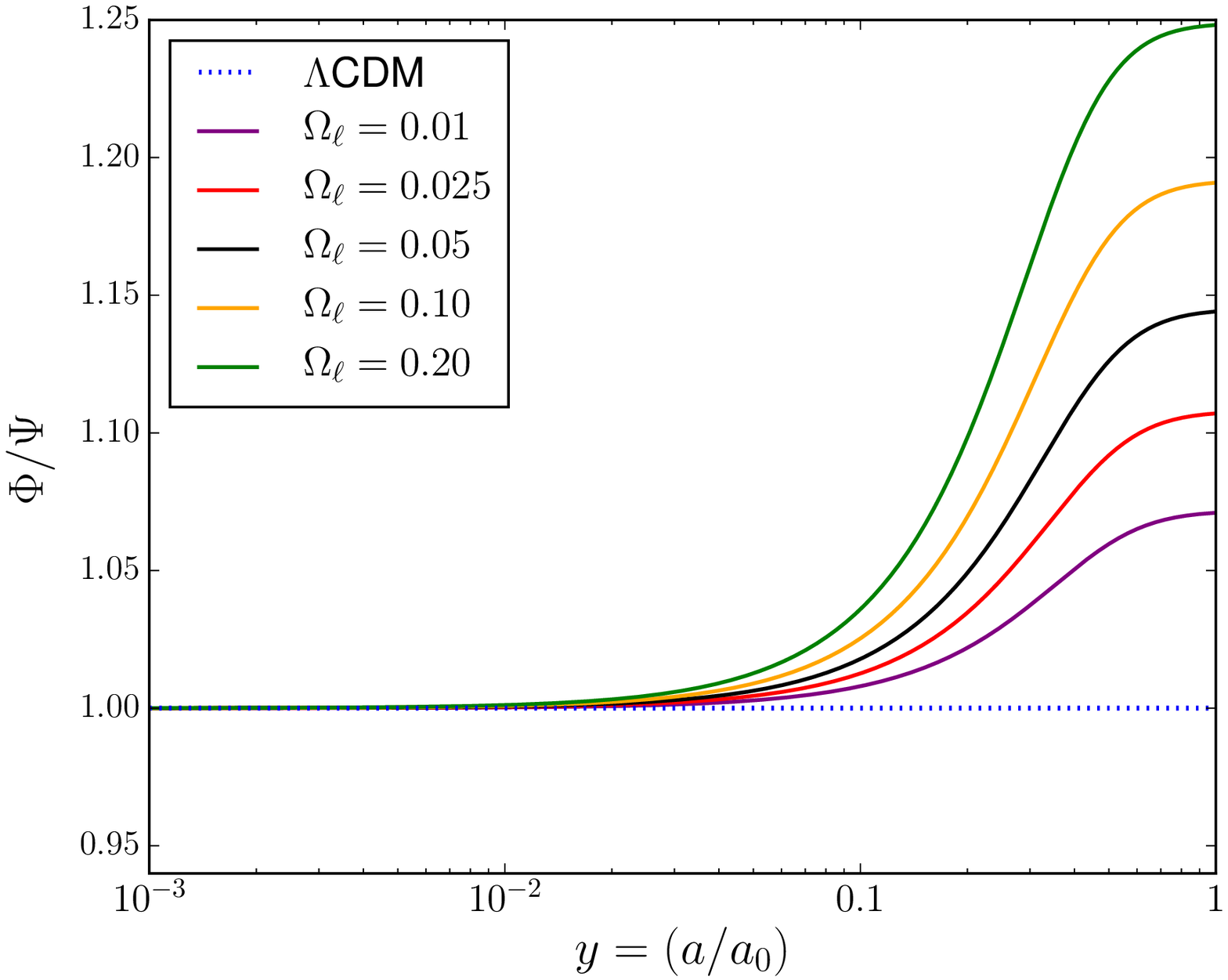}\label{fig:ratio_potentials}}
\subfigure[]{
\includegraphics[width=0.488\textwidth]{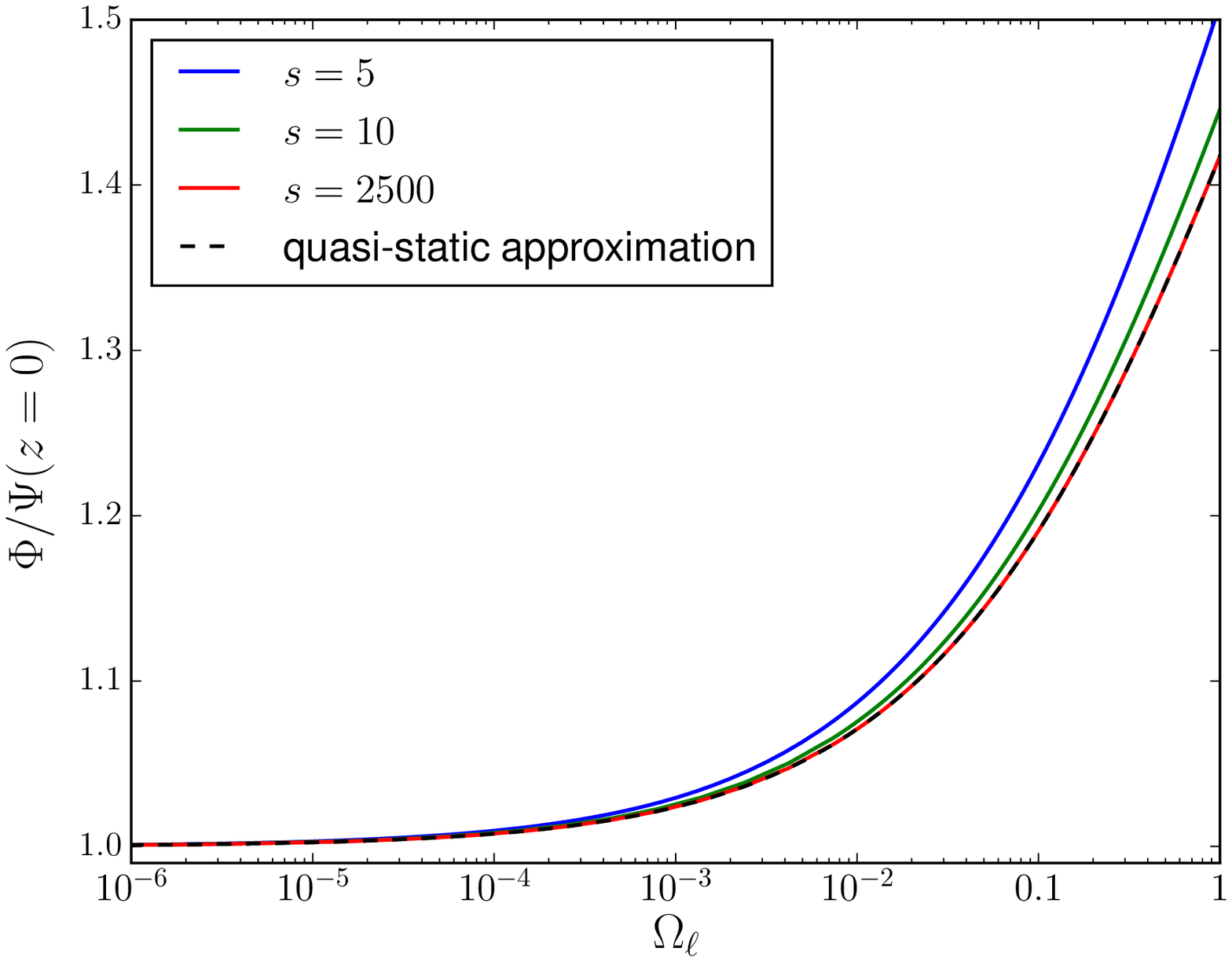}\label{fig:ratio_potentials1}}
\caption{{\bf(a):} The evolution of the ratio $\Phi/\Psi$ is shown for different values
of the brane parameter $\Omega_\ell$. Note that $\Phi/\Psi$ increases with $\Omega_\ell$,
where $\Omega_{\ell}$, defined in (\ref{beta}) and
(\ref{eq:cosmo_parameters}), depends on the ratio of the five- and four-dimensional
gravitational couplings. By contrast, $\Phi = \Psi$ in $\Lambda$CDM (dotted blue line).
Results are shown for the length scale $s\equiv k/a_0H_0 \simeq 2500$. {\bf(b):} The
current value of $\Phi/\Psi$ is shown as a function of $\Omega_\ell$ for different values
of the length scale $s\equiv k/a_0H_0$. Note that the quasi-static approximation provides
an excellent fit to the full analysis for $s \gtrsim 100$. } \label{fig:ratio_pot}
\end{figure}

The corresponding evolution of the gravitational potential $\Psi$ is shown in
figure~\ref{fig:Psi_mat}, and that of the relative difference $(\Psi - \Phi)/\Psi$ in
figure~\ref{fig:diff_mat}. Figure~\ref{fig:diff_ol} shows the evolution of $(\Psi -
\Phi)/\Psi$ for different values of $\Omega_\ell$, whereas figure~\ref{fig:diff_ol1}
shows the present value of $(\Psi - \Phi)/\Psi$ as a function of $\Omega_\ell$ for
different values of the wavenumber $s \equiv k/a_0H_0$. The corresponding results for the
ratio $\Phi/\Psi$ are shown in figure \ref{fig:ratio_pot}. These figures clearly
demonstrate the following features of our model:

\begin{enumerate}

\item The potentials $\Phi$ and $\Psi$ depart from the $\Lambda$CDM behaviour,
    $\Phi=\Psi$, at late times.

\item The departure from $\Lambda$CDM is more pronounced for larger values of the
    parameter $\Omega_\ell$, which is defined by (\ref{beta}) and
    (\ref{eq:cosmo_parameters}) and which depends on the ratio of the five- and
    four-dimensional gravitational couplings.

\item The dependence of $\Phi$ and $\Psi$ on scale is very weak, as shown in figures
    \ref{fig:diff_ol1} and \ref{fig:ratio_potentials1}. For large values of $s
    \gtrsim 100$, the results saturate and coincide with those of the quasi-static
    approximation. Hence, quasi-static approximation is able to reproduce the exact
    results for larger values of the wavenumber $s \equiv k/a_0H_0 \gtrsim 100$,
    including the scales relevant for structure formation, given in \eqref{s
    structure formation}. For very large spatial scales, $s < 100$, the quasi-static
    approximation becomes less accurate, consistent with figures \ref{fig:scale} and
    \ref{fig:scale1}.

\end{enumerate}

\begin{figure}[hbt]
\centering
\subfigure[]{
\includegraphics[width=0.488\textwidth]{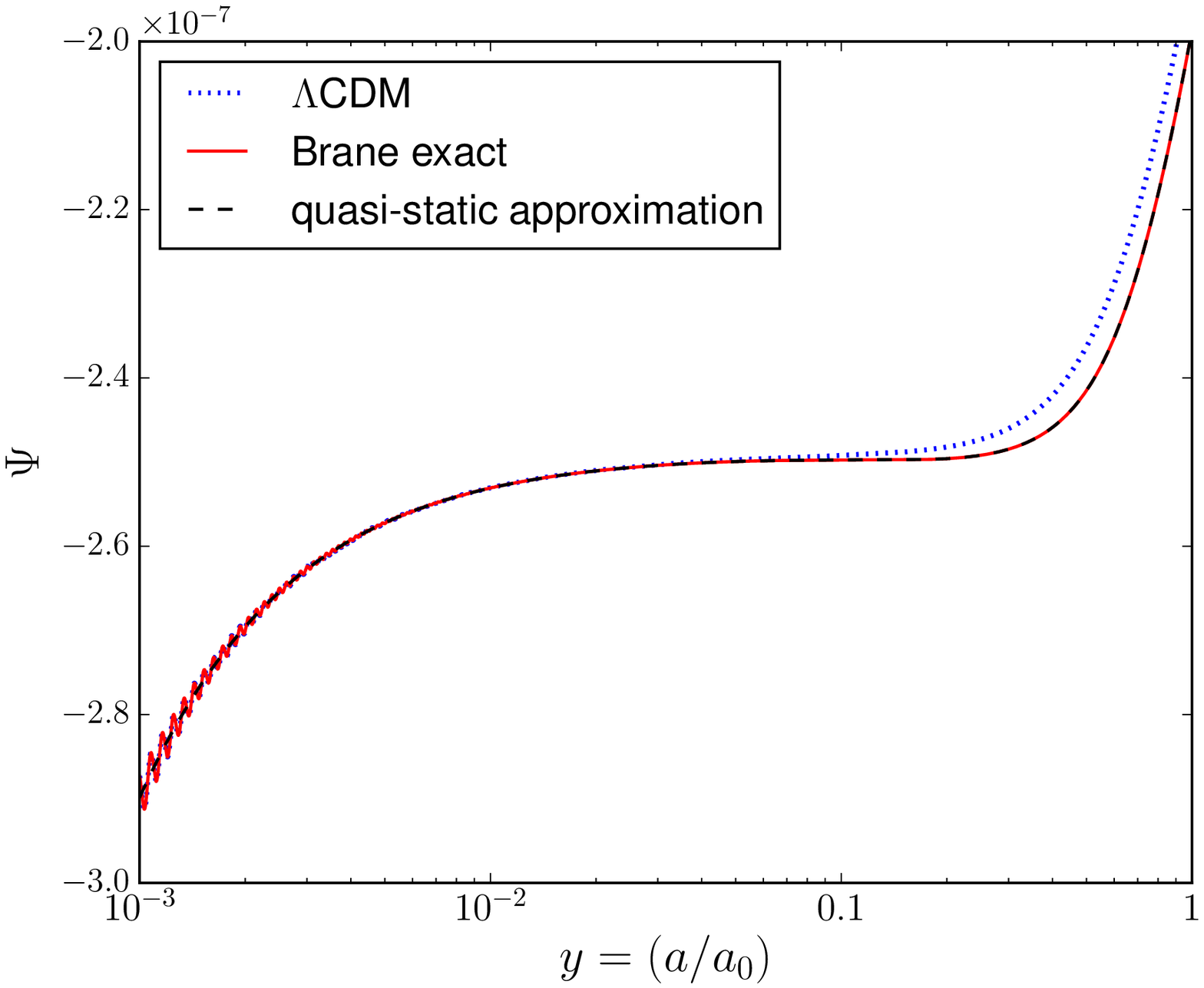}\label{fig:Psi_mat}}
\subfigure[]{
\includegraphics[width=0.488\textwidth]{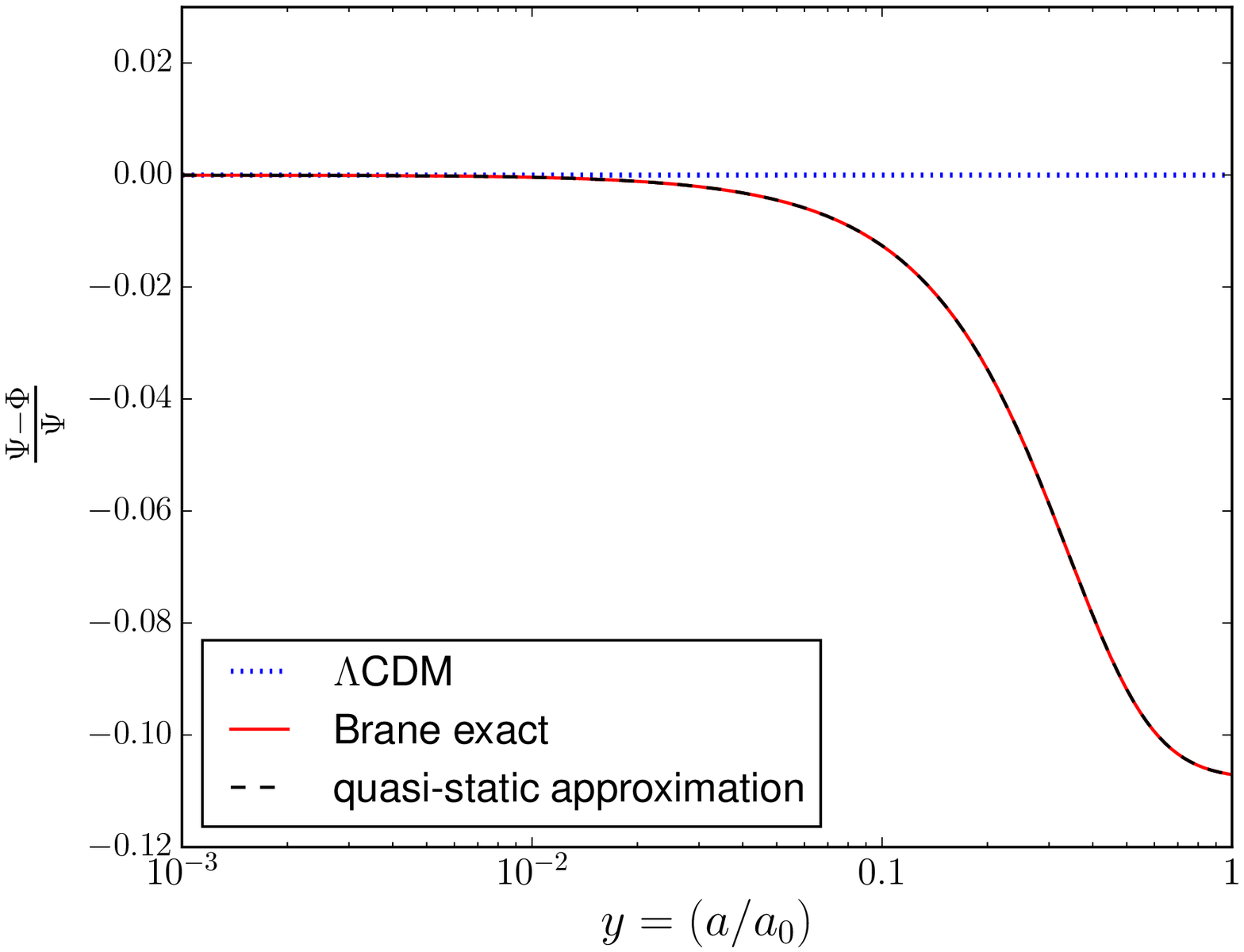}\label{fig:diff_mat}}
\caption{{\bf(a):} Evolution of the gravitational potential $\Psi$. Oscillations at early
times are due to perturbations in radiation and are sourced by the $\Delta_r$ term in
\eqref{Psi brane m-r}. Such oscillations are also  present in $\Lambda$CDM. {\bf(b):}
Evolution of the relative difference between the gravitational potentials $\Psi$ and
$\Phi$. The fractional difference $(\Psi - \Phi)/\Psi$ is very sensitive to the value of
$\Omega_{\ell}$ and marginally sensitive to the value of $s$, as illustrated in figures
\ref{fig:diff_ol} \& \ref{fig:diff_ol1}. Note the excellent accuracy of the quasi-static
approximation. Here we used $\Omega_{\ell}=0.025$ and $s\equiv k/a_0H_0 \simeq 2500$ for
the numerical illustrations. } \label{fig:pot}
\end{figure}

\iffalse
\begin{figure}[hbt]
\centering
\includegraphics[width=0.55\textwidth]{figures/Psi_mat.eps}
\caption{Evolution of the gravitational potential $\Psi$ for $\Omega_{\ell}=0.025$ and
$s\equiv k/a_0H_0 \simeq 2500$. Oscillations at early times are due to perturbations in
radiation and are sourced by the $\Delta_r$ term in \eqref{Psi brane m-r}. Such
oscillations are also  present in $\Lambda$CDM\@. } \label{fig:Psi_mat}
\end{figure}

\begin{figure}[htb]
\centering
\includegraphics[width=0.55\textwidth]{figures/r_PsimPhi_mat.eps}
\caption{Evolution of the relative difference between the gravitational potentials $\Psi$
and $\Phi$ is shown for $\Omega_{\ell}=0.025$ and $s\equiv k/a_0H_0 \simeq 2500$. Note
the excellent accuracy of the quasi-static approximation. The fractional difference
$(\Psi - \Phi)/\Psi$ is very sensitive to the value of $\Omega_{\ell}$ and marginally
sensitive to the value of $s$, as illustrated in figures \ref{fig:diff_ol} and
\ref{fig:diff_ol1}. } \label{fig:diff_mat}
\end{figure}
\fi

\begin{figure}[htb]
\centering
\subfigure[]{
\includegraphics[width=0.488\textwidth]{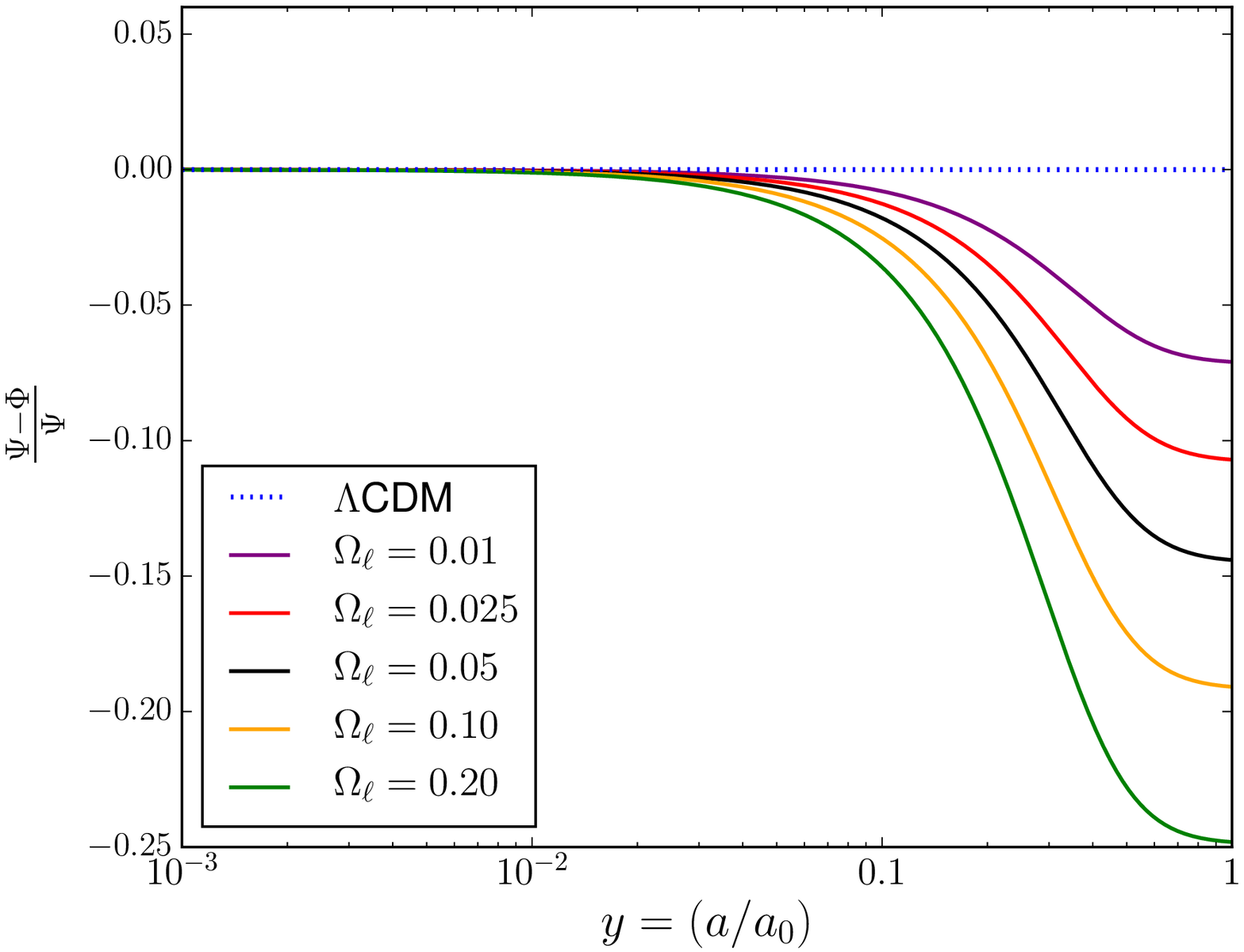}\label{fig:diff_ol}}
\subfigure[]{
\includegraphics[width=0.488\textwidth]{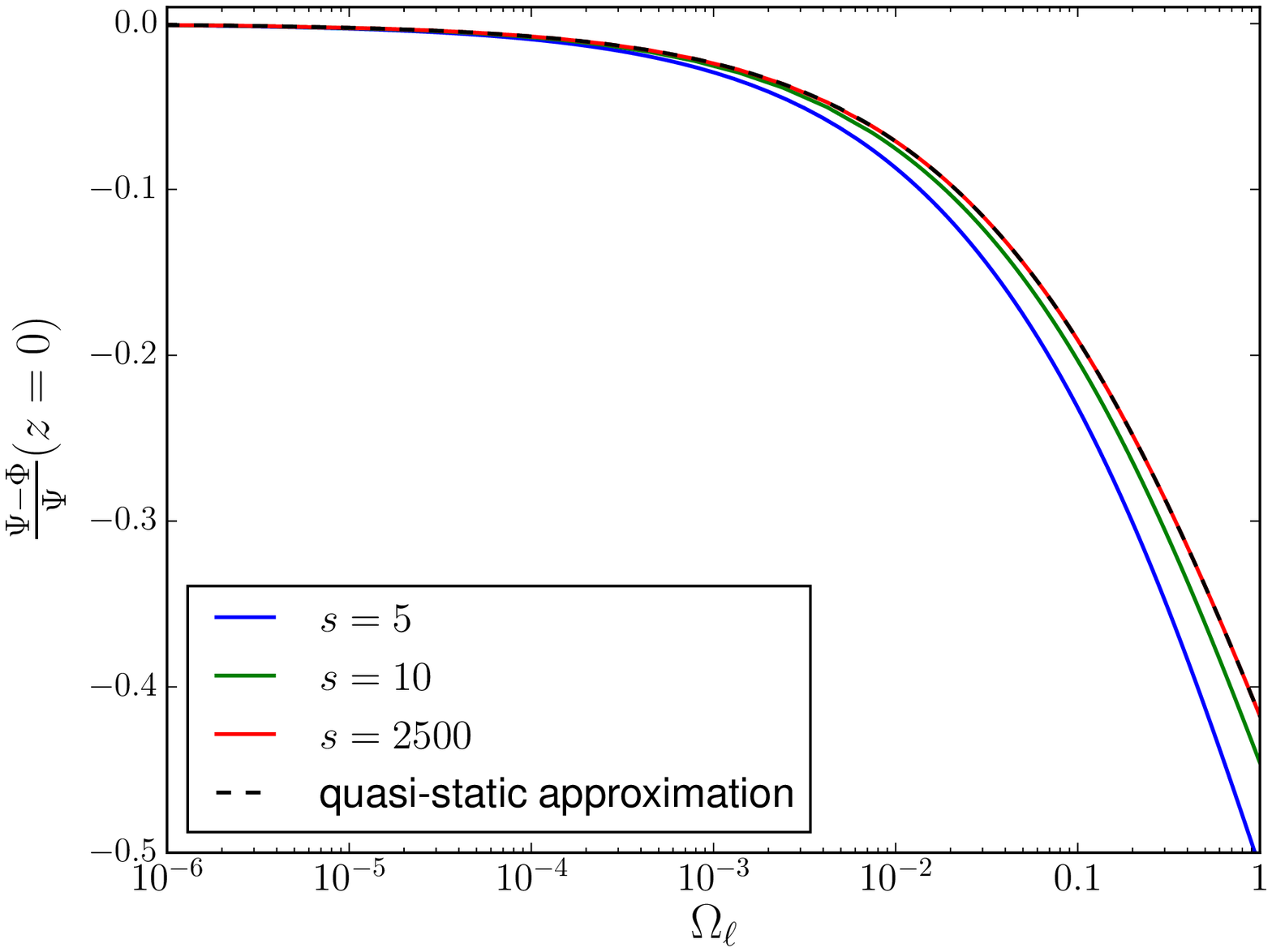}\label{fig:diff_ol1}}
\caption{{\bf(a):} The evolution of the relative difference between the gravitational
potentials is shown for various values of the brane parameter $\Omega_\ell$. Note that
the difference $\Psi - \Phi$ increases with the increase in $\Omega_\ell$, where
$\Omega_{\ell}$, defined by (\ref{beta}) and (\ref{eq:cosmo_parameters}),
depends on the ratio of the five- and four-dimensional gravitational couplings. Our
results are shown for the length scale $s\equiv k/a_0H_0 \simeq 2500$. {\bf(b):} The
relative difference between the gravitational potentials, evaluated at the present epoch,
is shown as a function of the brane parameter $\Omega_\ell$ for different values of the
wavenumber $s$. For very large spatial scales, $s < 100$, the fractional difference
between potentials deviates from that of the quasi-static approximation, given in
\eqref{eq:Psi_Phi_frac}. On the other hand, for $s \gtrsim 100$, the result saturates and
converges to the quasi-static approximation. } \label{fig:diff_ol_all}
\end{figure}

\iffalse
\begin{figure}[!htb]
\centering
\includegraphics[width=0.55\textwidth]{figures/r_PsimPhi_mat_ol.eps}
\caption{The evolution of the relative difference between the gravitational
potentials is shown as a function of the brane parameter $\Omega_\ell$.
Note that the difference $\Psi - \Phi$ increases with larger $\Omega_\ell$, where
$\Omega_{\ell}$, defined in
(\ref{beta}) \& (\ref{eq:cosmo_parameters}), describes the ratio of the five and four dimensional Planck mass.
Our results are shown for the length scale $s\equiv k/a_0H_0 \simeq 2500$.
} \label{fig:diff_ol}
\end{figure}

\begin{figure}[!htb]
\centering
\includegraphics[width=0.55\textwidth]{figures/rPsimPhi_ol_present.eps}
\caption{The relative difference between the gravitational
potentials, evaluated at the present epoch, is shown as a function of the brane parameter $\Omega_\ell$
 for different values of the wavenumber $s\equiv k/a_0H_0$. Our results converge to the
quasi-static approximation for $s\gtrsim 100$.
} \label{fig:diff_ol1}
\end{figure}
\fi

\iffalse
\begin{figure}[!htb]
\centering
\includegraphics[width=0.55\textwidth]{figures/PhibyPsi_ol.eps}
\caption{The evolution of $\Phi/\Psi$
is shown as a function of the brane parameter $\Omega_\ell$.
Note that $\Phi/\Psi$ increases with larger $\Omega_\ell$, where
$\Omega_{\ell}$, defined in
(\ref{beta}) \& (\ref{eq:cosmo_parameters}), describes the ratio of the five and four
dimensional Planck mass. By contrast $\Phi = \Psi$ in $\Lambda$CDM (dotted blue line).
Results are shown for the length scale $s\equiv k/a_0H_0 \simeq 2500$. }
\label{fig:ratio_potentials}
\end{figure}

\begin{figure}[!htb]
\centering
\includegraphics[width=0.55\textwidth]{figures/PhibyPsi_ol_present.eps}
\caption{The current value of $\Phi/\Psi$ is shown as a function of $\Omega_\ell$ for
different values of the length scale $s\equiv k/a_0H_0$. Note that the quasi-static
approximation provides an excellent fit to the full analysis for $s \gtrsim 100$. }
\label{fig:ratio_potentials1}
\end{figure}
\fi

\clearpage
%%%%%%%%%%%%%%%%%%%%%%%%%%%%%%%%%%%%%%%%%%%%%%%%%%%%%%%%%%%%%%%%%%%%%%%%%%%%%%%%%%%%%%%%%%5
\section{Conclusions}
\label{sec: conclusion}

We have investigated the evolution of perturbations on the normal branch of the induced
gravity braneworld, in which the brane is embedded in a flat bulk space-time. Of special
interest to us was the behavior of the bulk mode of perturbations, which is characterized
by non-zero (and possibly quite large) initial amplitude of effective Weyl fluid
perturbations.

Our approach to the problem, which is described in Sec.~\ref{sec: Scalar cosmological
perturbations on the brane}, goes beyond the quasi-static approximation and allows one to
study the behavior of perturbations starting from deep within the radiation-dominated
epoch, where the corresponding modes are super-Hubble. In Sec.~\ref{sec: Before the HG
crossing brane}, we established that perturbations of the Weyl fluid and those of matter
perturbations depend very weakly on time in the super-Hubble regime. This allows one to
set initial conditions during this early cosmological epoch, as described in
Sec.~\ref{sec: Setting initial conditions}.  We consider the initial perturbations of
matter and of the Weyl fluid as being statistically independent.  In this case, the
initial conditions describe two natural modes, which we refer to as the `brane mode' and
the `bulk mode', depending upon which perturbation vanishes initially. The
behavior of the {\em brane mode\/} on super-Hubble spatial scales is shown to be in good
agreement with the predictions of the scaling approximation \cite{Sawicki:2006jj,
Song:2007wd, Seahra:2010fj}. On sub-Hubble spatial scales, both methods converge to the
regime well described by the quasi-static approximation of \cite{Koyama:2005kd}.  The
{\em bulk mode\/} originates from perturbations with nonzero initial conditions in the
bulk; it was usually ignored in other approaches.

In Sec.~\ref{sec: After the HG crossing brane}, we established that Weyl-fluid
perturbations decrease with time in the bulk mode while growing in the brane mode after
the Hubble-radius crossing during radiation domination.  At the same time, perturbations
of radiation, as well as those of the gravitational potentials $\Phi$ and $\Psi$,
demonstrate the general-relativistic behavior, $\Phi \simeq \Psi$, in this regime.

At the beginning of the matter-dominated epoch, the contribution from the bulk mode
continues to decrease, as argued in Sec.~\ref{sec: Matter-dominated epoch
general-relativistic regime}. This explains the validity of the quasi-static
approximation, which was described in Sec.~\ref{sec: Koyama-Maartens approximation}.

All our results are confirmed by numerical integration of the exact system of equations
for pressureless matter, radiation and the Weyl fluid, which is performed in
Sec.~\ref{sec: Numerical integration}.  They are also in good agreement with
the five-dimensional numerical simulations of \cite{Cardoso:2007xc, Seahra:2010fj}.

Our main conclusion is that the presence of oscillatory terms in the evolution of
perturbations (like those described by equation \eqref{Delta matter domination g-r
solution}) are not likely to be of any significance, and the main effects of the
braneworld ansatz would be the almost self-similar ($\Omega_\ell\,$-dependent)
smooth deviation from general-relativistic behavior, shown in
figures~\ref{fig:ol-dep}, \ref{fig:Psi_mat}, \ref{fig:diff_mat}, \ref{fig:diff_ol}.

Our main results are summarized below.

\begin{enumerate}

\item Perturbations on the brane grow more rapidly than in the $\Lambda$CDM model at
    late times. This was illustrated in figure \ref{fig:ol-dep}. Departure from
    $\Lambda$CDM is more pronounced for larger values of the parameter $\Omega_\ell$,
    which is defined in (\ref{beta}) and (\ref{eq:cosmo_parameters}) and which
    depends on the ratio of the five- and four-dimensional gravitational couplings.
    (Note that the braneworld model under consideration passes to $\Lambda$CDM in the
    limit $\Omega_{\ell} \to 0$.)

\item Departure from $\Lambda$CDM is also reflected in the behaviour of the
    potentials $\Phi$ and $\Psi$ which follow the $\Lambda$CDM asymptote, $\Phi =
    \Psi$, only at early times. At late times, corresponding to $z \lesssim 50$, the
    difference between $\Phi$ and $\Psi$ becomes pronounced, and this effect is
    larger for larger values of $\Omega_\ell$.

\item The evolution of density perturbations and of the potentials $\Phi$ and $\Psi$
    displays a very weak dependence on length scale. This was illustrated in figures
    \ref{fig:scale} and \ref{fig:scale1} for the density contrast $\Delta_m$, and in
    figures \ref{fig:diff_ol1} and \ref{fig:ratio_potentials1} for the ratios
    $(\Psi-\Phi)/\Psi$ and $\Phi/\Psi$, respectively. These figures illustrate that
    our results for $\Delta_m$, $\Phi$ and $\Psi$ converge to those of the
    quasi-static approximation for larger values of the wavenumber $s \equiv k/a_0H_0
    \gtrsim 100$, which encompass the scales relevant for the large-scale structure
    formation, given in \eqref{s structure formation}.
\end{enumerate}

The results of this work will be compared with observations in a companion paper.

\section*{Acknowledgments}

S.~B. thanks the Council of Scientific and Industrial Research (CSIR), India, for
financial support as senior research fellow. A.~V., Y.~S. and V.~S. acknowledge support
from the India--Ukraine Bilateral Scientific Cooperation programme. The work of A.~V. and
Y.~S. is supported by the State Fund for Fundamental Research of Ukraine grant F64/45-2016.

%%%%%%%%%%%%%%%%%%%%%%%%%%%%%%%%%%%%%%%%%%%%%%%%%%%%%%%%%%%%%%%%%%%%%%%%%%%%%%%%%%%%%%%%%%%%%%%%%%%%%%%%%%

\end{document}